# Finding small separators in linear time via treewidth reduction[*]

Dániel Marx[†]    Barry O'Sullivan[‡]    Igor Razgon[§]

**Abstract**

We present a method for reducing the treewidth of a graph while preserving all of its minimal $s-t$ separators up to a certain fixed size $k$. This technique allows us to solve $s-t$ Cut and Multicut problems with various additional restrictions (e.g., the vertices being removed from the graph form an independent set or induce a connected graph) in linear time for every fixed number $k$ of removed vertices.

Our results have applications for problems that are not directly defined by separators, but the known solution methods depend on some variant of separation. For example, we can solve similarly restricted generalizations of Bipartization (delete at most $k$ vertices from $G$ to make it bipartite) in almost linear time for every fixed number $k$ of removed vertices. These results answer a number of open questions in the area of parameterized complexity. Furthermore, our technique turns out to be relevant for $(H,C,K)$- and $(H,C,\leq K)$-coloring problems as well, which are cardinality constrained variants of the classical $H$-coloring problem. We make progress in the classification of the parameterized complexity of these problems by identifying new cases that can be solved in almost linear time for every fixed cardinality bound.

## 1 Introduction

Finding cuts and separators is a classical topic in combinatorial optimization and in recent years there has been an increase of interest in the fixed-parameter tractability of such problems [7, 11, 24, 28, 30, 32, 50, 53, 54, 65]. Recall that a problem is *fixed-parameter tractable* (or FPT) with respect to a parameter $k$ if instances of size $n$ can be solved in time $f(k) \cdot n^{O(1)}$ for some computable function $f(k)$ depending only on the parameter $k$ of the instance [20, 25, 56]. In typical parameterized separation problems, the parameter $k$ is the size of the separator we are looking for, thus fixed-parameter tractability with respect to this parameter means that the combinatorial explosion is restricted to the size of the separator, but otherwise the running time depends polynomially on the size of the graph.

The main message of our paper is the following: the small $s-t$ separators live in a part of the graph that has bounded treewidth. Therefore, if a separation problem is FPT in bounded treewidth graphs, then it is FPT in general graphs as well. As there are general techniques for obtaining linear-time algorithms for problems on bounded-treewidth graphs (e.g., dynamic programming and Courcelle's Theorem), it follows that a surprisingly large number of generalized separation problems can be shown to be linear-time FPT with this approach (we say that a problem is *linear-time FPT* with parameter $k$ if it can be solved in time $f(k) \cdot n$ for some function $f$). For example, one of the consequences our general argument is a theorem stating that given a graph $G$, two terminal vertices $s,t$, and a parameter $k$, we can compute in a FPT-time a graph $G^*$ having the treewidth bounded by a function of $k$ while (roughly speaking) preserving all the inclusionwise minimal $s-t$ separators of size at most $k$. Therefore, combining this theorem with the well-known Courcelle's Theorem,

---

we obtain a powerful tool for finding $s-t$ separators obeying additional constraints expressible in monadic second order logic.

Algorithms for separation problems are often based on interesting mathematical properties of the problem. For example, the classical $s-t$ cut algorithm of Ford and Fulkerson is essentially based on the tight connection between maximum flows and minimum cuts. However, algorithms based on nice mathematical properties and connections are inherently fragile, and any slight generalization of the problem can break these connection and make the problem NP-hard (unless the generalization involves very special conditions, e.g., submodular functions). On the other hand, the main thesis of the paper is that the fixed-parameter tractability of separation problems has a highly robust theory: the technique of treewidth reduction presented in the paper allows us to show the fixed-parameter tractability of several generalizations with very little additional effort. As separation problems are crucial ingredients for solving other type of problems (e.g., bipartization), this robustness propagates into other problem areas as well.

## 1.1 Results

We demonstrate the power of the methodology with the following results.

- We prove that the MINIMUM STABLE $s-t$ CUT problem (Is there an independent set $S$ of size at most $k$ whose removal separates $s$ and $t$?) is fixed-parameter tractable and in fact can be solved in linear time for every fixed $k$. Our techniques allow us to prove various generalizations of this result very easily. First, instead of requiring that $S$ is independent, we can require that it induces a graph that belongs to a hereditary class $\mathcal{G}$ (hereditary means that if $G \in \mathcal{G}$, then every induced subgraph of $G$ is in $\mathcal{G}$ as well); the problem remains linear-time solvable for every fixed $k$. Second, in the MULTICUT problem a list of pairs of terminals are given $(s_1,t_1)$, ..., $(s_\ell,t_\ell)$ and $S$ is a set of at most $k$ vertices that induces a graph from $\mathcal{G}$ and separates $s_i$ from $t_i$ for every $i$. We show that this problem can be solved in linear time for every fixed $k$ and $\ell$ (i.e., linear-time FPT parameterized by $k$ and $\ell$), which is a very strong generalization of previous results [30, 50, 65]. Third, the results generalize to the MULTICUT-UNCUT problem, where two sets $T_1$, $T_2$ of pairs of terminals are given, and $S$ has to separate every pair of $T_1$ and *should not* separate any pair of $T_2$.

- We show that CONNECTED $s$-$t$ CUT (Is there a set of of at most $k$ vertices that induces a connected graph and whose removal separates terminals $s$ and $t$?) is linear-time FPT. The significance of the result is that at first sight this problem does not seem to be amenable to our techniques: connectivity is not a hereditary property, and therefore the solution is not necessarily a minimal $s$-$t$ separator. However, with some problem-specific ideas related to connectivity, we can extend our approach to handle such a requirement. This suggests that our technique might be applicable to a much wider range of cut problems than the hereditary problems described above.

- As a demonstration, we show that the EDGE-INDUCED VERTEX CUT (Is there a set of at most $k$ edges such that removal of their endpoints separates two given terminals $s$ and $t$?) is linear-time FPT, answering an open problem posed in 2007 by Samer [14]. The motivation behind this problem is described in [61]. While the reader might not be particularly interested in this exotic variant of $s-t$ cut, we believe that it nicely demonstrates the message of the paper. Slightly changing the definition of a well-understood cut problem usually makes the problem NP-hard and determining the parameterized complexity of such variants directly is by no means obvious and seems to require problem-specific ideas in each case. On the other hand, using our techniques, the fixed-parameter tractability of many such problems can be shown in a uniform way with very little effort. Let us mention (without proofs) three more variants that can be treated in a similar way: (1) separate $s$ and $t$ by deleting at most $k$ edges and at most $k$ vertices, (2) in a 2-colored graph, separate $s$ and $t$ by deleting at most $k$ black and at most $k$ white vertices, (3) in a $k$-colored graph, separate $s$ and $t$ by deleting one vertex from each color class.

- The BIPARTIZATION problem asks if a given graph $G$ can be made bipartite by deleting at most $k$ vertices. Reed et al. [60] showed that the problem is FPT and Kawarabayashi and Reed [44] proved that the problem is almost linear-time FPT, i.e., can be solved in time $f(k) \cdot n \cdot \alpha(n,n)$, where $\alpha$ is the inverse Ackermann function. We prove that the variant STABLE BIPARTIZATION (Is there an independent set of size *at most $k$* whose removal makes the graph bipartite?) is almost linear-time FPT, answering an open question posed by Fernau [14]. Furthermore, we prove that EXACT STABLE BIPARTIZATION (Is there an independent set of size *exactly $k$* whose removal makes the graph bipartite?) is also almost linear-time FPT, answering an open question posed in 2001 by Díaz et al. [16]. This latter result might be somewhat surprising, as finding an independent set of size exactly $k$ is W[1]-hard, and hence unlikely to be FPT. As in the case of $s-t$ cuts, we introduce the generalization $\mathcal{G}$-BIPARTIZATION, where the at most $k$ vertices of the solution have to induce a graph belonging to the class $\mathcal{G}$; we show that this problem is almost linear-time FPT whenever $\mathcal{G}$ is decidable and hereditary. We also study the analogous edge-deletion version $\mathcal{G}$-EDGEBIPARTIZATION and show it to be FPT if $\mathcal{G}$ is decidable and closed under taking subgraphs.

- The BIPARTITE CONTRACTION problem asks if a given graph $G$ can be made bipartite by the *contraction* of at most $k$ edges. Very recently, Heggernes et al. [38] showed that this problem is FPT by presenting a nontrivial problem-specific algorithm. We observe that a simple corollary of our results on $\mathcal{G}$-EDGEBIPARTIZATION immediately shows that BIPARTITE CONTRACTION is almost-linear time FPT.

- Finally, we analyze the constrained bipartization problems in a more general environment of $(H,C,\leq K)$-coloring [16], where the parameter is the maximum number of vertices mapped to $C$ in the homomorphism and prove that the problem is almost linear-time FPT if the graph $H \setminus C$ consists of two adjacent vertices without loops. There have been significant efforts in the literature to fully characterize the complexity (i.e., to prove dichotomy theorems) of various versions of $H$-coloring [8, 22, 23, 26, 33–35, 39, 40]. The version studied here was introduced in [16–19], where it was observed that this problem family contains several classical concrete problems as special case, including some significant open problems. Thus obtaining a full dichotomy would require breakthroughs in parameterized complexity. Our result removes one of the roadblocks towards this goal.

As the results listed above demonstrate, our method leads to the solution of several independent problems; it seems that the same combinatorial difficulty lies at the heart of these problems. Our technique manages to overcome this difficulty and it is expected to be of use for further problems of similar flavor. We would like to emphasize that while designing FPT-time algorithms for bounded-treewidth graphs and in particular the use of Courcelle's Theorem is a fairly standard method, we use this technique for problems where there is *no bound* on the treewidth in the input.

Various versions of (multiterminal) cut problems [11, 28, 32, 50] play a mysterious, not yet fully understood role in the fixed-parameter tractability of certain problems. Proving that BIPARTIZATION [60], DIRECTED FEEDBACK VERTEX SET [12], and ALMOST 2-SAT [58] are FPT answered longstanding open questions, and in each case the algorithm relies on a nonobvious use of separators. Furthermore, EDGE MULTICUT has been observed to be equivalent to FUZZY CLUSTER EDITING, a correlation clustering problem [1, 6, 15]. Thus aiming for a better understanding of separators in a parameterized setting seems to be a fruitful direction of research. The results of this paper extend our understanding of separators by showing that various additional constraints can be easily accommodated. It is important to point out that our algorithm is very different from previous parameterized algorithms for separation problems [7, 11, 28, 30, 32, 50, 54].

Those algorithms in the literature exploited certain nice properties of separators, and hence it seems very difficult to generalize them for the problems we consider here. On the other hand, our approach is very robust and, as demonstrated by our examples, it is able to handle many variants.

## 2 Treewidth reduction

The main combinatorial result of the paper is presented in this section. We start by introducing the main tools required to prove the result: the notions of treewidth and torso.

### 2.1 Treewidth, brambles, and monadic second order logic

A *tree decomposition* of a graph $G(V,E)$ is a pair $(T,\mathcal{B})$ in which $T(I,F)$ is a tree and $\mathcal{B} = \{B_i \mid i \in I\}$ is a family of subsets of $V(G)$ such that

1. $\bigcup_{i \in I} B_i = V$;
2. for each edge $e = (u,v) \in E$, there exists an $i \in I$ such that both $u$ and $v$ belong to $B_i$; and
3. for every $v \in V$, the set of nodes $\{i \in I \mid v \in B_i\}$ forms a connected subtree of $T$.

The *width* of the tree decomposition is the maximum size of a bag in $\mathcal{B}$ minus 1. The *treewidth* of a graph $G$, denoted by $\text{tw}(G)$, is the minimum width over all possible tree decompositions of $G$. For more background on the combinatorial and algorithmic consequences, the reader is referred to e.g., [5, 29]. A useful fact that we will use later on is that for every clique $K$ of $G$, there is a bag $B_i$ with $K \subseteq B_i$.

Treewidth has a dual characterization in terms of brambles [59, 62]. A *bramble* in a graph $G$ is a family of connected subgraphs of $G$ such that any two of these subgraphs either have a nonempty intersection or are joined by an edge. The *order* of a bramble is the least number of vertices required to cover all subgraphs in the bramble. The *bramble number* $\text{bn}(G)$ of a graph $G$ is the largest order of a bramble of $G$. Seymour and Thomas [62] proved that bramble number tightly characterizes treewidth:

**Theorem 2.1** (Seymour and Thomas [62]). *For every graph $G$, $\text{bn}(G) = \text{tw}(G) + 1$.*

Typically, the definition of treewidth is useful when we are trying to prove upper bounds, and brambles are useful when we are trying to prove lower bounds. Interestingly, in the current paper we use brambles to prove upper bounds on the treewidth. The reason for this is that we are relating the treewidth of different graphs appearing in our construction and want to show that if the resulting graph has large treewidth, then the earlier graphs have large treewidth as well.

The algorithmic importance of treewidth comes from the fact that a large number of NP-hard problems can be solved in linear time if we have a bound on the treewidth of the input graph. Most of these algorithms use a bottom-up dynamic programming approach, which generalizes dynamic programming on trees. Courcelle's Theorem [13] (see also [20, Section 6.5], [29]) gives a powerful way of quickly showing that a problem is linear-time solvable on bounded treewidth graphs. Sentences in *Monadic Second Order Logic of Graphs* (MSO) contain quantifiers, logical connectives ($\neg$, $\vee$, and $\wedge$), vertex variables, vertex set variables, binary relations $\in$ and $=$, and the atomic formula $E(u,v)$ expressing that $u$ and $v$ are adjacent. If a graph property can be described in this language, then this description can be turned into an algorithm:

**Theorem 2.2** (Courcelle [13]). *If a graph property can be described as a formula $\phi$ in the Monadic Second Order Logic of Graphs, then it can be recognized in time $f_\phi(\text{tw}(G)) \cdot (|E(G)| + |V(G)|)$ if a given graph $G$ has this property.*

Theorem 2.2 can be extended to labeled graphs, where the sentence contains additional atomic formulas $P_i(x)$ meaning that vertex $x$ has label $i$. We can implement labels on the edges by additional atomic formulas $E_i(x,y)$ with the meaning that there is an edge of label $i$ connecting vertices $x$ and $y$. We informally call these labels as "colors" and talk e.g., about colored graphs with black and white vertices and red and blue edges. Most of the results in the paper go through for graphs colored with fixed constant number of colors: the colors do not play a role in graph-theoretic properties (such as separation, treewidth, etc.) and requirements on colors can be easily accommodated in MSO formulas.

Constructing an MSO formula for a given graph problem is usually a straightforward, but somewhat lengthy exercise. Thus when we use Theorem 2.2, the construction of the formula is relegated to the appendix.

## 2.2 Separators

Two slightly different notions of separation will be used in the paper:

**Definition 2.3.** *We say that a set $S$ of vertices separates sets of vertices $A$ and $B$ if no component of $G \setminus S$ contains vertices from both $A \setminus S$ and $B \setminus S$. If $s$ and $t$ are two distinct vertices of $G$, then an $s-t$ separator is a set $S$ of vertices disjoint from $\{s,t\}$ such that $s$ and $t$ are in different components of $G \setminus S$.*

Thus if we say that $S$ separates $A$ and $B$, then we do not require that $S$ is disjoint from $A$ and $B$. In particular, if $S$ separates $A$ and $B$, then $A \cap B \subseteq S$.

We say that an $s-t$ separator $S$ is *minimum* if there is no $s-t$ separator $S'$ with $|S'| < |S|$. We say that an $s-t$ separator $S$ is *(inclusionwise) minimal* if there is no $s-t$ separator $S'$ with $S' \subset S$.

If $X$ is a set of vertices, we denote by $N_G(X)$ the set of those vertices in $V(G) \setminus X$ that are adjacent to at least one vertex of $X$. (We omit the subscript $G$ if it clear from the context.) We use the folklore result that all the minimum cuts can be covered by a sequence of noncrossing minimum cuts: there exists a sequence $X_1 \subset \cdots \subset X_q$ such that every $N(X_i)$ is a minimum $s-t$ separator and every vertex that appears in a minimum separator is covered by one of the $N(X_i)$'s. The existence of these sets can be proved by a simple application of the uncrossing technique. We present a different proof here (related to [57]) that allows us to find such a sequence in linear time. Strictly speaking, it is not possible to construct the sets $X_1, \ldots, X_q$ in linear time, as their total size could be quadratic. However, it is sufficient to produce the differences $X_{i+1} \setminus X_i$, as they contain all the information in the sequence.

**Lemma 2.4.** *Let $s,t$ be two vertices in graph $G$ such that the minimum size of an $s-t$ separator is $\ell > 0$. Then there is a collection $\mathcal{X} = \{X_1, \ldots X_q\}$ of sets where $\{s\} \subseteq X_i \subseteq V(G) \setminus (\{t\} \cup N(\{t\}))$ $(1 \leq i \leq q)$, such that*

1. *$X_1 \subset X_2 \subset \cdots \subset X_q$,*
2. *$|N(X_i)| = \ell$ for every $1 \leq i \leq q$, and*
3. *every $s-t$ separator of size $\ell$ is fully contained in $\bigcup_{i=1}^{q} N(X_i)$.*

*Furthermore, there is an $O(\ell(|V(G)| + |E(G)|))$ time algorithm that produces the sets $X_1$, $X_1 \setminus X_2$, ..., $X_q \setminus X_{q-1}$ corresponding to such a collection $\mathcal{X}$.*

*Proof.* Let us construct a directed network $D$ the following way. There are $2|V(G)|$ nodes in $D$: for every $v \in V(G)$, there are two nodes $v_1$, $v_2$, there is an arc $\overrightarrow{v_1v_2}$ with capacity 1, and there is an arc $\overrightarrow{v_2v_1}$ with infinite capacity. For every edge $xy \in E(G)$, we add two arcs $\overrightarrow{x_2y_1}$ and $\overrightarrow{y_2x_1}$ with infinite capacity.

For $Y \subseteq V(D)$, let $\Delta_D^+(Y)$ be the set of edges leaving $Y$ in $D$. We say that $F \subseteq E(D)$ is an $s_2 \to t_1$ *cut*, if there is no path from $s_2$ to $t_1$ in $D \setminus F$. It is clear that a set $S \subseteq V(G) \setminus \{s,t\}$ is an $s-t$ separator if and only if the corresponding set $\{\overrightarrow{v_1v_2} \mid v \in S\}$ of $|S|$ arcs of $D$ form an $s_2 \to t_1$ cut. Therefore, if we can find a sequence $\{s_2\} \subseteq Y_1 \subset Y_2 \subset \cdots \subset Y_q \subseteq V(D) \setminus \{t_1\}$ such that the capacity of $\Delta_D^+(Y_i)$ is $\ell$ for every $1 \leq i \leq q$ and every $s_2 \to t_1$ cut of weight $\ell$ is fully contained in $\bigcup_{i=1}^q \Delta_D^+(Y_i)$, then we can obtain the required sequence by defining $X_i$ to contain those vertices $v$ for which $v_1, v_2 \in Y_i$. It is easy to observe that in this case $v \in N(X_i)$ if and only if the corresponding arc $\overrightarrow{v_1 v_2}$ is in $\Delta_D^+(Y_i)$

Let us run $\ell$ rounds of the Ford-Fulkerson algorithm on the network $D$ to find a maximum $s_2 \to t_1$ flow and let $D'$ be the residual graph (recall that the residual graph $D'$ contains an arc $\overrightarrow{uv}$ if and only if either $D$ contains an unsaturated arc $\overrightarrow{uv}$ in or an arc $\overrightarrow{vu}$ with nonzero flow). Let $C_1, \ldots, C_q$ be a topological ordering of the strongly connected components of $D'$ (i.e. $i < j$ whenever there is a path from $C_i$ to $C_j$). As the value of the maximum flow is exactly $\ell > 0$, there is no $s_2 \to t_1$ path in the residual graph $D'$, but there has to be a $t_1 \to s_2$ path. Therefore, if $t_1$ is in $C_x$ and $s_2$ is in $C_y$, then $x < y$. For every $x < i \leq y$, let $Y_i := \bigcup_{j=i}^q C_j$. We claim that the capacity of $\Delta_D^+(Y_i)$ is $\ell$. By the definition of $Y_i$, no arc leaves $Y_i$ in the residual graph $D'$, hence every edge leaving $Y_i$ in $D$ is saturated and no flow enters $Y_i$. As $s_2 \in C_y \subseteq Y_i$ and $t_1 \in C_x \subseteq V(G) \setminus Y_i$, this is only possible if the capacity of $\Delta_D^+(Y_i)$ is exactly $\ell$.

What remains to be shown is that every arc contained in an $s_2 \to t_1$ cut of weight $\ell$ is covered by one of the $\Delta_D^+(Y_i)$'s. Let $F$ be an $s_2 \to t_1$ cut of weight $\ell$. Let $Y$ be the set of nodes reachable from $s_2$ in $D \setminus F$; as $F$ is a minimum cut, it is clear that $\Delta_D^+(Y) = F$. Consider an arc $\overrightarrow{ab} \in F$. Since this arc is in a minimum cut, it is saturated by the flow, hence there is an arc $\overrightarrow{ba}$ entering $Y$ in the residual graph $D'$. We claim that this arc does not appear in any cycle of $D'$. If it appears in a cycle, then there is an arc $\overrightarrow{cd}$ of $D'$ leaving $Y$. However, such an arc cannot exist, as every arc leaving $Y$ in $D$ is saturated and no flow enters $Y$. Thus $a$ and $b$ are in two different strongly connected components $C_{i_a}$ and $C_{i_b}$ for some $i_b < i_a$. Since there is flow going from $s_2$ to $a$, there is an $a \to s_2$ path in the residual graph $D'$, and hence $i_a \leq y$. Similarly, as there is flow going from $b$ to $t_1$, there is a $t_1 \to b$ path in $D'$ and $i_b \geq x$. Thus we have that $x \leq i_b < i_a \leq y$, hence $Y_{i_a}$ is well-defined, and $\overrightarrow{ab}$ of $D$ is contained in $\Delta_D^+(Y_{i_a})$. □

## 2.3 Torso

If we are interested in those separators of graph $G$ that are fully contained in a subset $C$ of vertices, then we can replace the neighborhood of each component in $G \setminus C$ by a clique, as there is no way to disconnect these vertices with separators fully contained in $C$. The notion of torso and Proposition 2.7 below formalize this concept.

**Definition 2.5.** *Let $G$ be a graph and $C \subseteq V(G)$. The graph* $\text{torso}(G,C)$ *has vertex set $C$ and vertices $a, b \in C$ are connected by an edge if $\{a,b\} \in E(G)$ or there is a path $P$ in $G$ connecting $a$ and $b$ whose internal vertices are not in $C$.*

We state without proof some easy consequences of the definition:

**Proposition 2.6.** *Let $G$ be a graph.*

1. *For sets $C_1 \subseteq C_2 \subseteq V(G)$, we have $\text{torso}(\text{torso}(G,C_2),C_1) = \text{torso}(G,C_1)$.*
2. *For sets $C, S \subseteq V(G)$, we have that $\text{torso}(G \setminus S, C \setminus S)$ is a subgraph of $\text{torso}(G,C) \setminus S$.*

The following proposition states that, roughly speaking, the operation of taking the torso preserves the separators that are contained in $C$.

**Proposition 2.7.** *Let $C_1 \subseteq C_2$ be two subsets of vertices in $G$ and let $a, b \in C_1$ two vertices. A set $S \subseteq C_1$ separates $a$ and $b$ in $\text{torso}(G,C_1)$ if and only if $S$ separates these vertices in $\text{torso}(G,C_2)$. In particular, by setting $C_2 = V(G)$, we get that $S \subseteq C_1$ separates $a$ and $b$ in $\text{torso}(G,C_1)$ if and only if it separates them in $G$.*

*Proof.* Assume first that $C_2 = V(G)$, that is, $\text{torso}(G,C_2) = G$. Let $P$ be a path connecting $a$ and $b$ in $G$ and suppose that $P$ is disjoint from a set $S$. The path $P$ contains vertices from $C_1$ and from $V(G) \setminus C_1$. If $u,v \in C_1$ are two vertices such that every vertex of $P$ between $u$ and $v$ is from $V(G) \setminus C_1$, then by definition there is an edge $uv$ in $\text{torso}(G,C_1)$. Using these edges, we can modify $P$ to obtain a path $P'$ that connects $a$ and $b$ in $\text{torso}(G,C_1)$ and avoids $S$.

Conversely, suppose that $P$ is a path connecting $a$ and $b$ in $\text{torso}(G,C_1)$ and it avoids $S \subseteq C_1$. If $P$ uses an edge $uv$ that is not present in $G$, then this means that there is a path connecting $u$ and $v$ whose internal vertices are not in $C_1$. Using these paths, we can modify $P$ to obtain a path $P'$ that uses only the edges of $G$. Since $S \subseteq C_1$, the new vertices on the path are not in $S$, i.e., $P'$ avoids $S$ as well.

For the general statement observe that it follows from the previous paragraph that $S \subseteq C_1$ separates $a$ and $b$ in $\text{torso}(\text{torso}(G,C_2),C_1)$ if and only if it separates $a$ and $b$ in $\text{torso}(G,C_2)$. Now the statement of the proposition immediately follows from Prop. 2.6(1). □

We show that if we have a treewidth bound on $\text{torso}(G,C_i)$ for every $i$, then these bounds add up for the union of the $C_i$'s. The characterization of treewidth using bramble number will be convenient for the proof of this statement.

**Lemma 2.8.** *Let $G$ be a graph and $C_1, \ldots, C_r$ be subsets of $V(G)$ and let $C := \bigcup_{i=1}^{r} C_i$. We have $\text{bn}(\text{torso}(G,C)) \leq \sum_{i=1}^{r} \text{bn}(\text{torso}(G,C_i))$.*

*Proof.* It is sufficient to prove the lemma in the case $C = V(G)$: otherwise, we prove the statement for the graph $G' := \text{torso}(G,C)$ (note that $\text{torso}(G,C_i) = \text{torso}(G',C_i)$ by Prop. 2.6(1)).

Let $\mathcal{B}$ be a bramble of $G$ having order $\text{bn}(G)$. For every $1 \leq i \leq r$, Let $\mathcal{B}_i = \{B \cap C_i \mid B \in \mathcal{B}, B \cap C_i \neq \emptyset\}$. We claim that $\mathcal{B}_i$ is a bramble of $\text{torso}(G,C_i)$. Let us show first that $B \cap C_i \in \mathcal{B}_i$ is connected. Consider two vertices $x, y \in B \cap C_i$. Since $B \in \mathcal{B}$ is connected, there is an $x - y$ path $P$ in $G$ such that every vertex of $P$ is in $B$. As in the proof of Prop. 2.7, we can replace the subpaths of $P$ that leave $C_i$ by edges of $\text{torso}(G,C_i)$ to obtain an $x - y$ path $P'$ such that every vertex of $P'$ is in $B \cap C_i$. This proves that $B \cap C_i$ is connected.

Let us show now that the sets in $\mathcal{B}_i$ pairwise touch. Consider two sets $B_1 \cap C_i, B_2 \cap C_i \in \mathcal{B}_i$. We know that $B_1$ and $B_2$ touch in $G$, thus there are vertices $x \in B_1$, $y \in B_2$ such that either $x = y$ or $x$ and $y$ are adjacent in $G$. If $x, y \in C_i$, then it is clear that $B_1 \cap C_i$ and $B_2 \cap C_i$ touch in $\text{torso}(G,C_i)$. If $x$ or $y$ is not in $C_i$, then both $x$ and $y$ are in $K \cup N_G(K)$ for some component $K$ of $G \setminus C_i$. As $B_1 \cap C_i, B_2 \cap C_i \neq \emptyset$, both $B_1$ and $B_2$ have to intersect $N_G(K)$. By the definition of torso, $N_G(K)$ induces a clique in $\text{torso}(G,C_i)$, thus $B_1 \cap C_i$ and $B_1 \cap C_i$ touch in $\text{torso}(G,C_i)$.

Let $X_i$ be a cover of $\mathcal{B}_i$ having order $\text{bn}(\text{torso}(G_i))$. To complete the proof of the lemma, we observe that $\bigcup_{i=1}^{r} X_i$ is a cover of $\mathcal{B}$. Indeed, as $C = V(G)$, every $B \in \mathcal{B}$ intersects some $C_i$, and therefore $X_i$ intersects $B \cap C_i$. □

The following lemma will be used for the inductive proof of the treewidth reduction result. This time, the definition of treewidth using tree decompositions will be more useful in the proof.

**Lemma 2.9.** *Let $C' \subseteq V(G)$ be a set of vertices and let $R_1, \ldots, R_r$ be components of $G \setminus C'$. For every $1 \leq i \leq r$, let $C'_i \subseteq R_i$ be subsets and let $C'' := C' \cup \bigcup_{i=1}^{r} C'_i$. Then we have*

$$\text{tw}(\text{torso}(G,C'')) \leq \text{tw}(\text{torso}(G,C')) + \max_{i=1}^{r} \text{tw}(\text{torso}(G[R_i],C'_i)) + 1$$

$$\text{bn}(\text{torso}(G,C'')) \leq \text{bn}(\text{torso}(G,C')) + \max_{i=1}^{r} \text{bn}(\text{torso}(G[R_i],C'_i)).$$

*Proof.* Let $T$ be a tree decomposition of $\text{torso}(G,C')$ having width at most $w_1$, and let $T_i$ be a tree decomposition of $\text{torso}(G[R_i],C'_i)$ having width at most $w_2$. Let $N_i \subseteq C'$ be the neighborhood of $R_i$ in $G$. Since $N_i$ induces a clique in $\text{torso}(G,C')$, we have $|N_i| \leq w_1 + 1$ and there is a bag $B_i$ of $T$ containing $N_i$. Let us modify $T_i$ by including $N_i$ in every bag and then let us join $T$ and $T_i$ by connecting an arbitrary bag of $T_i$ with $B_i$. By performing this step for every $1 \le i \le r$, we get a tree decomposition having width at most $w_1 + w_2 + 1$. To show that it is indeed a tree decomposition of $\mathrm{torso}(G, C'')$, consider two vertices $x, y \in C''$ that are adjacent in $\mathrm{torso}(G, C'')$, i.e., there is a path $P$ between them whose internal vertices are outside $C''$.

1. If $x, y \in C'$, then they are adjacent in $\mathrm{torso}(G, C')$ as well (since the internal vertices of $P$ are outside $C'' \supseteq C'$) and hence they appear together in a bag of $T$.
2. If $x, y \in C'_i$, then all the internal vertices of $P$ are in $R_i$. Thus they are adjacent in $\mathrm{torso}(G[R_i], C'_i)$ and hence they appear together in a bag of $T_i$.
3. If $x \in C'$ and $y \in C'_i$, then $x \in N_i$ and every bag of $T_i$ containing $y$ was extended with the $N_i$.

$\square$

A simple consequence of Lemma 2.9 is that adding a constant number of vertices to $C$ increases the treewidth of the torso only by a constant:

**Corollary 2.10.** *For every graph $G$, sets $C, X \subseteq V(G)$, we have $\mathrm{tw}(\mathrm{torso}(G, C \cup X)) \le \mathrm{tw}(\mathrm{torso}(G, C)) + |X|$.*

On the other hand, removing even a single vertex from $C$ can increase treewidth arbitrarily. For example, let $G$ be a star with a central vertex $v$ and $n$ leaves. Then $\mathrm{tw}(\mathrm{torso}(G, V(G))) = \mathrm{tw}(G) = 1$, while $\mathrm{torso}(G, V(G) \setminus \{v\})$ is a clique on $n$ vertices, hence it has treewidth $n - 1$.

## 2.4 Treewidth bound for minimal $s-t$ separators

If the minimum size of an $s-t$ separator is $\ell$, then the *excess* of an $s-t$ separator $S$ is $|S| - \ell$ (which is always nonnegative). Note that if $s$ and $t$ are adjacent, then no $s-t$ separator exists, and in this case we say that the minimum size of an $s-t$ separator is $\infty$. The aim of this section is to show that, for every $k$, we can construct a set $C'$ covering all the minimal $s-t$ separators of size at most $k$ such that $\mathrm{torso}(G, C')$ has treewidth bounded by a function of $k$. Equivalently, we can require that $C'$ covers every minimal $s-t$ separator of excess of at most $e := k - \ell$, where $\ell$ is the minimum size of an $s-t$ separator. Observe that it is crucial that we want to cover only inclusionwise minimal separators: every vertex (other than $s$ and $t$) is part of some $s-t$ separator having excess 1.

Lemma 2.4 shows that the union $C$ of all minimum $s-t$ separators can be covered by a chain of minimum $s-t$ separators. It is not difficult to see that this chain can be used to define a tree decomposition (in fact, a path decomposition) of $\mathrm{torso}(G, C)$. This observation solves the problem for $e = 0$. For the general case, we use induction on $e$.

**Lemma 2.11.** *Let $s, t$ be two vertices of graph $G$ and let $\ell$ be the minimum size of an $s-t$ separator. For some $e \ge 0$, let $C$ be the union of all minimal $s-t$ separators having* excess *at most $e$ (i.e. having size at most $k = \ell + e$). Then there is an $f(\ell, e) \cdot (|E(G)| + |V(G)|)$ time algorithm that returns a set $C' \supseteq C$ disjoint from $\{s, t\}$ such that $\mathrm{bn}(\mathrm{torso}(G, C')) \le g(\ell, e)$, for some functions $f$ and $g$ depending only on $\ell$ and $e$.*

*Proof.* We prove the lemma by induction on $e$. Consider the collection $\mathcal{X}$ of Lemma 2.4 and define $S_i := N(X_i)$ for $1 \le i \le q$. For the sake of uniformity, we define $X_0 := \emptyset$, $X_{q+1} := V(G) \setminus \{t\}$, $S_0 := \{s\}$, $S_{q+1} := \{t\}$. For $1 \le i \le q+1$, let $L_i := X_i \setminus (X_{i-1} \cup S_{i-1})$. See Figure 2.4 for an intuitive illustration of these sets. Observe that $L_i$'s are pairwise disjoint. Furthermore, for $1 \le i \le q+1$ and two disjoint *non-empty* subsets $A, B$ of $S_i \cup S_{i-1}$, we define $G_{i,A,B}$ to be the graph obtained from $G[L_i \cup A \cup B]$ by contracting the set $A$ to a vertex $a$ and the set $B$ to a vertex $b$. The key observation that makes it possible to use induction is that if $C$ includes a vertex of some $L_i$, then $e > 0$.

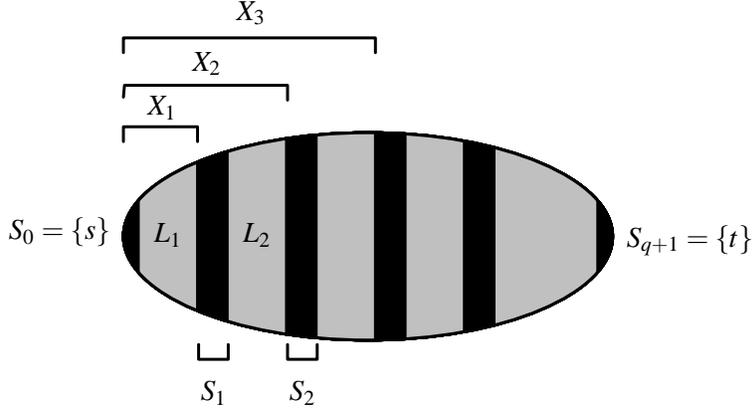

Figure 1: Schematic illustration of the first few sets $S_i$ and $L_i$ in the proof of Lemma 2.11. The illustration is simplified, e.g., it does not take into account that $S_{i-1}$ and $S_i$ are not necessarily disjoint.

**Claim 2.12.** *If a vertex $v \in L_i$ is in C, then there are disjoint non-empty subsets $A, B$ of $S_i \cup S_{i-1}$ such that $v$ is part of a minimal $a - b$ separator $K_2$ in $G_{i,A,B}$ of size at most $k$ (recall that $k = \ell + e$) and excess at most $e - 1$.*

*Proof.* By definition of $C$, there is a minimal $s - t$ separator $K$ of size at most $k$ that contains $v$. Let $K_1 := K \setminus L_i$ and $K_2 := K \cap L_i$. Partition $(S_i \cup S_{i-1}) \setminus K$ into the set $A$ of vertices reachable from $s$ in $G \setminus K$ and the set $B$ of vertices non-reachable from $s$ in $G \setminus K$. Let us observe that both $A$ and $B$ are non-empty. Indeed, due to the minimality of $K$, $G$ has a path $P$ from $s$ to $t$ such that $V(P) \cap K = \{v\} \subseteq L_i$. By selection of $v$, $S_{i-1}$ separates $v$ from $s$ and $S_i$ separates $v$ from $t$. Therefore, at least one vertex $u$ of $S_{i-1}$ occurs in $P$ before $v$ and at least one vertex $w$ of $S_i$ occurs in $P$ after $v$. The prefix of $P$ ending at $u$ and suffix of $P$ starting at $w$ are both subpaths in $G \setminus K$. It follows that $u$ is reachable from $s$ in $G \setminus K$, i.e. belongs to $A$ and that $w$ is reachable from $t$ in $G \setminus K$, hence non-reachable from $s$ and thus belongs to $B$.

To see that $K_2$ is an $a - b$ separator in $G_{i,A,B}$, suppose that there is a path $P$ connecting $a$ and $b$ in $G_{i,A,B}$ avoiding $K_2$. Then there is a corresponding path $P'$ in $G$ connecting a vertex of $A$ and a vertex of $B$. Path $P'$ is disjoint from $K_1$ (since it contains vertices of $L_i$ and $(S_i \cup S_{i-1}) \setminus K$ only) and from $K_2$ (by construction). Thus a vertex of $B$ is reachable from $s$ in $G \setminus K$, a contradiction.

To see that $K_2$ is a minimal $a - b$ separator, suppose that there is a vertex $u \in K_2$ such that $K_2 \setminus \{u\}$ is also an $a - b$ separator in $G_{i,A,B}$. Since $K$ is minimal, there is an $s - t$ path $P$ in $G \setminus (K \setminus u)$, which has to pass through $u$. Arguing as when we proved that $A$ and $B$ are non-empty, we observe that $P$ includes vertices of both $A$ and $B$, hence we can consider a minimal subpath $P'$ of $P$ between a vertex $a' \in A$ and a vertex $b' \in B$. We claim that all the internal vertices of $P'$ belong to $L_i$. Indeed, due to the minimality of $P'$, an internal vertex of $P'$ can belong either to $L_i$ or to $V(G) \setminus (K_1 \cup L_i \cup S_{i-1} \cup S_i)$. If all the internal vertices of $P'$ are from the latter set then there is a path from $a'$ to $b'$ in $G \setminus (K_1 \cup L_i)$ and hence in $G \setminus (K_1 \cup K_2)$ in contradiction to $b' \in B$. If $P'$ contains internal vertices of both sets then $G$ has an edge $\{u, w\}$ where $u \in L_i$ while $w \in V(G) \setminus (K_1 \cup L_i \cup S_{i-1} \cup S_i)$. But this is impossible since $S_{i-1} \cup S_i$ separates $L_i$ from the rest of the graph. Thus it follows that indeed all the internal vertices of $P'$ belong to $L_i$. Consequently, $P'$ corresponds to a path in $G_{i,A,B}$ from $a$ to $b$ that avoids $K_2 \setminus u$, a contradiction that proves the minimality of $K_2$.

Finally, we show that $K_2$ has excess at most $e-1$. Let $K_2'$ be a minimum $a-b$ separator in $G_{i,A,B}$. Observe that $K_1 \cup K_2'$ is an $s-t$ separator in $G$. Indeed, consider a path $P$ from $s$ to $t$ in $G \setminus (K_1 \cup K_2')$. It necessarily contains a vertex $u \in K_2$, hence arguing as in the previous paragraph we notice that $P$ includes vertices of both $A$ and $B$. Considering a minimal subpath $P'$ of $P$ between a vertex $a' \in A$ and $b' \in B$ we observe, analogously to the previous paragraph that all the internal vertices of this path belong to $L_i$. Hence this path correspond to a path between $a$ and $b$ in $G_{i,A,B}$. It follows that $P'$, and hence $P$, includes a vertex of $K_2'$, a contradiction showing that $K_1 \cup K_2'$ is indeed an $s-t$ separator in $G$. Due to the minimality of $K_2$, $K_2' \neq \emptyset$. Thus $K_1 \cup K_2'$ contains at least one vertex from $L_i$, implying that $K_1 \cup K_2'$ is not a minimum $s-t$ separator in $G$. Thus $|K_2| - |K_2'| = (|K_1| + |K_2|) - (|K_1| + |K_2'|) < k - \ell = e$, as required. This completes the proof of Claim 2.12. □

Now we define $C'$. Let $C_0 := \bigcup_{i=1}^{q} S_i$ (note that $s, t \notin C_0$). In the case $e = 0$, we can set $C' = C_0$. It is easy to see that $\text{tw}(\text{torso}(G, C_0))) \leq 2\ell - 1$: the bags $S_1 \cup S_2, S_2 \cup S_3, \ldots, S_{q-1} \cup S_q$ define a tree decomposition of width at most $2\ell - 1$.

Assume now that $e > 0$. For every $1 \leq i \leq q + 1$ and disjoint non-empty subsets $A, B$ of $S_i \cup S_{i-1}$, the induction assumption implies that there exists a set $C'_{i,A,B} \subseteq L_i$ such that $\text{bn}(\text{torso}(G_{i,A,B}, C'_{i,A,B})) \leq g(\ell, e-1)$ and $C'_{i,A,B}$ contains every inclusionwise minimal $a-b$ separator of size at most $k$ and excess at most $e-1$ in $G_{i,A,B}$. We define $C'$ as the union of $C_0$ and all sets $C'_{i,A,B}$ as above. Observe that $C'$ satisfies the requirement that any vertex $v$ participating in a minimal $s-t$ separator of size at most $k$ indeed belongs to $C'$: every such separator of size $\ell$ is contained in $C_0$, and if the separator has size strictly greater than $\ell$, then Claim 2.12 implies that $v$ is contained in some $C'_{i,A,B}$.

We would like to use Lemma 2.9 to show that the bramble number of $\text{torso}(G, C')$ can be bounded by a function $g(\ell, e)$. Each component of $G \setminus C_0$ is fully contained in some $L_i$. Let $C'_i$ be the union of the at most $3^{2\ell}$ sets $C'_{i,A,B}$, for nonempty disjoint subsets $A, B \subseteq S_i \cup S_{i-1}$. As $G[L_i] = G_{i,A,B} \setminus \{a, b\}$ and $C'_{i,A,B}$ is disjoint from $\{a, b\}$, we have that $\text{torso}(G[L_i], C'_{i,A,B}) = \text{torso}(G_{i,A,B}, C'_{i,A,B})$. Therefore, by Lemma 2.8, the bramble number of $\text{torso}(G[L_i], C'_i)$ is at most $3^{2\ell} \cdot g(k, e-1)$. It follows that we have the same bound on the bramble number of $\text{torso}(G[R], C' \cap R)$ for every component $R$ of $G \setminus C_0$. Recall that the treewidth of $\text{torso}(G, C_0)$ is at most $2\ell - 1$, hence $\text{bn}(\text{torso}(G, C_0)) \leq 2\ell$. Therefore, $\text{bn}(\text{torso}(G, C')) \leq 2\ell + 3^{2\ell} \cdot g(k, e-1)$ holds by Lemma 2.9.

We conclude the proof by showing that the above set $C'$ can be constructed in time $f(\ell, e) \cdot (|E(G)| + |V(G)|)$ for an appropriate function $f(\ell, e)$. We prove this statement by induction on $e$. The sets $X_i$ can be computed as shown in the proof of Lemma 2.4. Then the sets $S_i$ can be obtained in the first paragraph of the proof of the present lemma. Their union results in $C_0$ which is $C'$ for $e = 0$. Thus for $e = 0$, $C'$ can be computed in time $O(k(|V(G)| + |E(G)|))$. Now assume that $e > 0$. For each $i$ such that $1 \leq i \leq q + 1$ and $|L_i| > 0$, the algorithm explores all possible disjoint subsets $A$ and $B$ of $S_i \cup S_{i-1}$. Let $m_i$ be the number of edges of $G[L_i]$. Observe that the number of edges of $G_{i,A,B}$ is at most $m_i + 2|L_i|$: the degree of the two extra vertices $a$ and $b$ is at most $|L_i|$. For the given choice, we check if the size of a minimum $a-b$ separator of $G_{i,A,B}$ is at most $k$ (observe that this can be done in time $O(k(m_i + 2|L_i|))$ by $k$ rounds of the Ford-Fulkerson algorithm) and if yes, compute the set $C'_{i,A,B}$ recursively. Thus the number of steps required to handle layer $i$ (not including the recursive calls) is $O(3^{2\ell} \cdot k \cdot (m_i + 2|L_i|))$. By the induction assumption, each of the at most $3^{2\ell}$ recursive calls takes at most $f(\ell, e-1) \cdot (m_i + 2|L_i|)$ steps. Therefore, the overall running time of

computing $C'$ is

$$O(|E(G)|) + \sum_{i=1}^{q+1} \left(3^{2\ell}O(k(m_i + 2|L_i|)) + 3^{2\ell}f(\ell, e-1)(m_i + 2|L_i|)\right)$$
$$\leq O(|E(G)|) + 3^{2\ell}O(k(E(G) + 2|V(G)|)) + 3^{2\ell}f(\ell, e-1)(E(G) + 2|V(G)|)$$
$$\leq f(\ell, e)(|E(G)| + |V(G)|),$$

for an appropriate choice of $f(\ell,e)$ (the first inequality follows from the fact that the $L_i$'s are disjoint, and hence $\sum_{i=1}^{q+1}|L_i| \leq |V(G)|$ and $\sum_{i=1}^{q+1}m_i \leq 2|E(G)|$). □

**Remark 2.13.** The set $C'$ given by Lemma 2.11 is disjoint from $\{s,t\}$, thus torso$(G, C')$ does not contain $s$ or $t$. However, by Corollary 2.10, extending $C'$ with $s$ and $t$ increases treewidth only by 2.

**Remark 2.14.** The recursion $g(\ell, e) := 2\ell + 3^{2\ell} \cdot g(\ell, e-1))$ implies that $g(\ell, e)$ is $2^{O(e\ell)}$, i.e., the treewidth/bramble number bound is exponential in $\ell$ and $e$. It is an obvious question whether it is possible to improve this dependence to polynomially bounded. However, a simple example shows that the function $g(\ell, e)$ has to be exponential. Let $G$ be the $n$-dimensional hypercube and let $s$ and $t$ be opposite vertices. The size of the minimum $s-t$ separator is $\ell := n$. We claim that every vertex $v$ of the hypercube (other than $s$ and $t$) is part of a minimal $s-t$ separator of size at most $n(n-1)$. To see this, let $P$ be a shortest path connecting $s$ and $v$. Let $P' = P - v$ be the subpath of $P$ connecting $s$ with a neighbor $v'$ of $v$. Let $S$ be the neighborhood of $P'$; clearly $S$ is an $s-t$ separator and $v \in S$. However, $S \setminus v$ is not an $s-t$ separator: the path $P$ is not blocked by $S \setminus v$ as $S \setminus v$ does not contain any vertex farther from $s$ than $v$. Since $P'$ has at most $n-1$ vertices and every vertex has degree $n$, we have $|S| \leq n(n-1)$. Thus $v$ (and every other vertex) is part of a minimal separator of size at most $n(n-1)$. Hence if we set $\ell := n$ and $e := n(n-1) - n$, then $C$ contains every vertex of the hypercube (except $s$ and $t$). The treewidth of an $n$-dimensional hypercube is $\Omega(2^n/\sqrt{n})$ [10], which is also a lower bound on $g(\ell, e)$.

Lemma 2.11 together with Prop. 2.7 show that if we want to find an $s-t$ separator of size at most $k$ satisfying some additional constraints, then it is sufficient to find such an $s-t$ separator in the bounded-treewidth graph torso$(G, C' \cup \{s,t\})$, which can be done using standard techniques. However, there is a minor technical detail here. The graph torso$(G, C' \cup \{s,t\})$ can contain edges not originally present in $G$. Therefore, it is possible that some $S \subseteq C'$ satisfies the required property (say, inducing an independent set) in $G$, but not in torso$(G, C' \cup \{s,t\})$ (or vice versa). This problem can be solved by marking the new edges as "special": for example, when solving the problem using Courcelle's Theorem, we can assume that the input structure contains a binary predicate distinguishing these new edges. Intuitively, we can say that the original edges of the graph have color "black," the new edges introduced by the torso operation are "red," and we are looking for a separator where the black edges form a certain patters. In Theorem 2.15, we follow a different approach: we modify the graph such that every minimal $s-t$ separator $S$ of size at most $k$ induces the same graph in the original graph $G$ and in the new bounded-treewidth graph. This theorem states our main combinatorial tool in a form that will be very convenient to use for applications where we require that $S$ induces a certain type of graphs.

**Theorem 2.15.** [The treewidth reduction theorem] *Let $G$ be a graph, $T \subseteq V(G)$, and let $k$ be an integer. Let $C$ be the set of all vertices of $G$ participating in a minimal $s-t$ separator of size at most $k$ for some $s,t \in T$. For every fixed $k$ and $|T|$, there is a linear-time algorithm that computes a graph $G^*$ having the following properties:*

1. *$C \cup T \subseteq V(G^*)$*
2. *For every $s,t \in T$, a set $K \subseteq V(G^*)$ with $|K| \leq k$ is a minimal $s-t$ separator of $G^*$ if and only if $K \subseteq C \cup T$ and $K$ is a minimal $s-t$ separator of $G$.*

3. *The treewidth of $G^*$ is at most $h(k,|T|)$ for some function h.*
4. *$G^*[C\cup T]$ is isomorphic to $G[C\cup T]$.*

*Proof.* For every $s,t \in T$ that can be separated by the removal of at most $k$ vertices, the algorithm of Lemma 2.11 computes a set $C'_{s,t}$ containing all the minimal $s-t$ separators of size at most $k$. By Lemma 2.8, if $C'$ is the union of these $\binom{|T|}{2}$ sets, then $\text{torso}(G,C')$ has treewidth bounded by a function of $k$ and $|T|$. By Corollary 2.10, we have similar bound on the treewidth of $G' = \text{torso}(G,C'\cup T)$. Note that $G'$ satisfies all the requirements of the theorem except the last one: two vertices of $C'$ non-adjacent in $G$ may become adjacent in $G'$ (see Definition 2.5). To fix this problem we subdivide each edge $\{u,v\}$ of $G'$ such that $\{u,v\} \notin E(G)$ with a new vertex, and, to avoid selection of this vertex into a cut, we split it into $k+1$ copies. In other words, for each edge $\{u,v\} \in E(G') \setminus E(G)$ we introduce $k+1$ new vertices $w_1,\ldots,w_{k+1}$ and replace $\{u,v\}$ by the set of edges $\{\{u,w_1\}\ldots\{u,w_{k+1}\},\{w_1,v\},\ldots,\{w_{k+1},v\}\}$. Let $G^*$ be the resulting graph. It is not hard to check that $G^*$ satisfies all the properties of the present theorem. □

**Remark 2.16.** The treewidth of $G^*$ may be larger than the treewidth of $G$. We use the phrase "treewidth reduction" in the sense that the treewidth of $G^*$ is bounded by a function of $k$ and $|T|$, while the treewidth of $G$ is unbounded in general.

## 3 Constrained separation problems

We present a set of results in this section that give linear-time algorithms for vertex cut problems where the cut has to have a certain property, for example, it induces a graph that belongs to a class $\mathcal{G}$.

### 3.1 Hereditary graph classes

Let $\mathcal{G}$ be a class of graphs. Given a graph $G$, vertices $s$, $t$, and parameter $k$, the $\mathcal{G}$-MINCUT problem asks if $G$ has an $s-t$ separator $C$ of size at most $k$ such that $G[C] \in \mathcal{G}$. Suppose that $\mathcal{G}$ is *hereditary,* i.e. for every $G \in \mathcal{G}$ and $X \subseteq V(G)$, we have $G[X] \in \mathcal{G}$. In this case, whenever there is an $s-t$ separator $K$ of size at most $k$ that induces a member of $\mathcal{G}$, then there is a *minimal $s-t$* separator of size at most $k$ that induces a member of $\mathcal{G}$: if $K' \subset K$ is a minimal $s-t$ separator, then $G[K] \in \mathcal{G}$ implies $G[K'] \in \mathcal{G}$. Therefore, if we construct (in linear time) the graph $G^*$ of Theorem 2.15 for $S = \{s,t\}$, then $G$ has an $s-t$ separator of size at most $k$ that induces a member of $\mathcal{G}$ if and only if $G^*$ has such a separator. This means that we can solve the problem on graph $G^*$, whose treewidth is bounded by a function of $k$. Courcelle's Theorem provides a very easy way to prove that certain problems are linear-time solvable on graphs of bounded treewidth. Using this result, it is a routine exercise to show that for every fixed bound $k$ on the separator size and fixed bound $w$ on the treewidth of the graph, $\mathcal{G}$-MINCUT is linear-time solvable: all we need to do is to construct an appropriate sentence in monadic second order logic expressing that there is a set of at most $k$ vertices that form an $s-t$ separator and induces a member of $\mathcal{G}$ (see Appendix A.1 for details). Note that the class $\mathcal{G}$ contains a finite number of graphs having at most $k$ vertices, and if $\mathcal{G}$ is decidable, then they can be enumerated in time depending only on $k$. Therefore, we obtain our first basic result:

**Theorem 3.1.** *Assume that $\mathcal{G}$ is decidable and hereditary. Then the $\mathcal{G}$-MINCUT problem can be solved in time $f_{\mathcal{G}}(k)(|E(G)|+|E(H)|)$.*

Theorem 3.1 allows us to answer a fairly natural open question in the area of parameterized complexity. In particular, let $\mathcal{G}^0$ be the class of all graphs without edges. Then $\mathcal{G}^0$-MINCUT is the MINIMUM STABLE $s-t$ CUT problem whose fixed-parameter tractability has been posed as an open question by Kanj [43]. Clearly, $\mathcal{G}^0$ is hereditary and hence the $\mathcal{G}^0$-MINCUT is FPT.

**Corollary 3.2.** MINIMUM STABLE $s-t$ CUT *is linear-time FPT.*

Theorem 3.1 can be used to decide if there is an $s-t$ separator of size *at most k* having a certain property, but cannot be used if we are looking for $s-t$ separators of size *exactly k*. We show (with a very easy argument) that some of these problems actually become hard if the size is required to be exactly $k$.

**Theorem 3.3.** *It is* W[1]-*hard (parameterized by k) to decide if G has an $s-t$ separator that is an independent set of size exactly k.*

*Proof.* Let $G$ be a graph and let $G'$ be a graph obtained from $G$ by adding two isolated vertices $s$ and $t$. As every set not containing $s$ and $t$ is an $s-t$ separator, $G$ has an independent set of size exactly $k$ if and only if $G'$ has an independent $s-t$ separator of size exactly $k$. Since it is W[1]-hard to check the existence of an independent set of size exactly $k$, it follows that it is also W[1]-hard to check existence of an independent $s-t$ separator of size exactly $k$. □

The hardness of checking the existence of a separator of size exactly $k$ that is a clique or a dominating set can be proven similarly.

We remark that Theorem 3.1 remains true for graphs having a fixed finite number of colors (the finite number of colors ensures that $\mathcal{G}$ contains only a finite number of graphs with at most $k$ vertices). Therefore, we can solve problems such as finding an $s-t$ separator having at most $k$ red vertices and at most $k$ blue vertices.

## 3.2 Edge-induced vertex cuts

Samer and Szeider [61] introduced the notion of *edge-induced vertex-cut* and the corresponding computational problem: given a graph $G$ and two vertices $s$ and $t$, the task is to find out if there is an $s-t$ separator that can be covered by $k$ edges. Intuitively, we want to separate $s$ and $t$ by removing the *endpoints* of at most $k$ edges (but of course we are not allowed to remove $s$ and $t$ themselves). It remained an open question in [61] whether this problem is FPT. Samer reposted this problem as an open question in [14]. Using Theorem 3.1, we answer this question positively. For this purpose, we introduce $\mathcal{G}_k$, the class of graphs where the number of vertices minus the size of the maximum matching is at most $k$, observe that this class is hereditary, and show that $(G,s,t,k)$ is a yes-instance of the *edge-induced vertex-cut* problem if and only if $(G,s,t,2k)$ is a yes-instance of the $\mathcal{G}_k$-MINCUT problem. Then we apply Theorem 3.1 to get the following corollary.

**Corollary 3.4.** *The* EDGE-INDUCED VERTEX-CUT *problem is linear-time solvable for every fixed k.*

*Proof.* Let $\mathcal{G}_k$ contain those graphs where the number of vertices minus the size of the maximum matching is at most $k$. It is not hard to observe that $\mathcal{G}_k$ is hereditary by noticing that for any $H \in \mathcal{G}_k$ and $v \in V(H)$ the difference between the number of vertices and the size of maximum matching does not increase by removal of $v$. It follows therefore from Theorem 3.1 that $\mathcal{G}_k$-MINCUT is FPT.

We will show that $\mathcal{G}_k$-MINCUT with parameter $2k$ is equivalent to the problem of finding out whether $s$ can be separated from $t$ by removal of a set $S$ that can be extended to the union of at most $k$ edges.

Assume that $(G,s,t,2k)$ is a yes-instance of the $\mathcal{G}_k$-MINCUT problem and let $S$ be an $s-t$ separator of size at most $2k$ such that $G[S] \in \mathcal{G}_k$. Since $\mathcal{G}_k$ is hereditary, we may assume that $S$ is a minimal $s-t$ separator. Let $M$ be a maximum matching of $G[S]$. Then, by definition of $\mathcal{G}_k$, we have $|S| - |M| \leq k$ or, in other words, $|S| - 2|M| \leq k - |M|$. The $2|M|$ vertices of $G[S]$ incident to the matching are covered by $|M|$ edges. The remaining at most $k - |M|$ vertices can be covered by selecting an edge of $G$ incident to each of them (due to the minimality of $S$, it does not contain isolated vertices). Thus $s$ and $t$ may be separated by removal of a set that can be covered by at most $k$ edges. Conversely, assume that $s$ and $t$ can be separated by removal of set $S$ of vertices that can be extended to the union of at most $k$ edges of $G$. Clearly $|S| \leq 2k$.

It is not hard to observe that the size of the smallest set of edges covering $S$ equals the size of the maximum matching $|M|$ of $G[S]$ plus $|S| - 2|M|$ edges for the vertices not covered by the matching. By definition of $S$, $|M| + |S| - 2|M| \leq k$. It follows that $G[S] \in \mathcal{G}_k$. Thus, $(G, s, t, 2k)$ is a yes-instance of the $\mathcal{G}_k$-MINCUT problem. □

## 3.3 Connected vertex cuts

The *stable cut* and *edge-induced vertex-cut* problems can be solved by direct application of Theorem 3.1. The following problem is an example where the required property is not hereditary, but the problem can be still handled with a slight extension of our framework. We say that an $s-t$ separator $S$ in graph $G$ is a *connected $s-t$ separator* if $G[S]$ is a connected graph. The main difficulty in finding a connected $s-t$ separator of size at most $k$ is that we cannot restrict our search to minimal $s-t$ separators: every connected $s-t$ separator $S$ contains a minimal $s-t$ separator $S' \subseteq S$, but there is no guarantee that $G[S']$ is connected as well. Therefore, we cannot assume that the solution $S$ is fully contained in the set $C'$ produced by Lemma 2.11.

The right way to look at the problem of finding a connected $s-t$ separator is that we have to find a minimal $s-t$ separator $S'$ that can be extended to a connected set $S$ of size at most $k$. Let us call a set *k-connectable* if it can be extended to a connected set of size at most $k$. Deciding if a given $S'$ is $k$-connectable is essentially a Steiner Tree problem: we are looking for a tree having at most $k$ vertices (or equivalently, at most $k-1$ edges) containing $S'$. Given a set $X$ of terminals in an edge-weighted graph, the classical algorithm of Dreyfus and Wagner [21] finds in time $3^{|X|} \cdot n^{O(1)}$ a minimum weight tree that contains $X$. Recently, this has been improved to $2^{|X|} \cdot n^{O(1)}$ for unweighted graphs [4, 48, 55]. However, neither of these algorithms is linear time. We need the following result, which shows that for every fixed *size k*, a tree of at most $k$ vertices containing $X$ can be found in linear time (if exists). It can be proved by a simple modification of the algorithm of Dreyfus and Wagner: for example, it is sufficient to solve recurrence (12) in [48] up to $i = k$.

**Lemma 3.5.** *Let $G$ be a graph, $X \subseteq V(G)$, and $k$ an integer. There is an $O(3^k(|V(G)| + |E(G)|))$ time algorithm that finds a tree containing $X$ and having at most $k$ vertices, if such a tree exists.*

The following lemma contains the main idea of our algorithm for finding connected $s-t$ separators. By our observations above, it is sufficient to look for a minimal $s-t$ separator that is $k$-connectable. Lemma 2.11 gives us a set $C'$ that contains every such minimal $s-t$ separator $S$, but it is possible that $S$ cannot be extended to a connected set of size at most $k$ inside $C'$. We show that by considering a slightly larger set $C''$, we can ensure that every $k$-connectable set in $C'$ can be extended to a connected set in $C''$.

**Lemma 3.6.** *Let $G$ be a graph with two vertices $s$ and $t$, and let $k$ be an integer. There is a set $C'' \subseteq V(G)$ such that $\mathrm{tw}(\mathrm{torso}(G, C''))$ is bounded by a constant depending only on $k$ and the following holds: whenever $G$ has a connected $s-t$ separator $S$ of size at most $k$, $G$ also has a connected separator $S'$ of size at most $k$ such that $S' \subseteq C''$. Furthermore, such a set $C''$ can be found in time $f(k)(|E(G)| + |V(G)|)$.*

*Proof.* Let us use the algorithm of Lemma 2.11 to obtain a set $C' \subseteq V(G)$ containing every minimal $s-t$ separator of size at most $k$. Let $K_1, \ldots, K_q$ be the components of $G \setminus C'$ and let $N_i$ be the neighborhood of $K_i$ in $C'$. As $N_i$ induces a clique in $\mathrm{torso}(G, C')$, we have that $|N_i| \leq \mathrm{tw}(\mathrm{torso}(G, C')) + 1$. For every $1 \leq i \leq q$ and every nonempty subset $X \subseteq N_i$ of size at most $k$, let us use the algorithm of Lemma 3.5 to check if there is a connected set $T_{i,X}$ of size at most $k$ such that $X \subseteq T_{i,X} \subseteq K_i \cup N_i$, and if so, let $T_{i,X}$ be such a set of minimum size; otherwise, let $T_{i,X} = \emptyset$. Let $T_i = \bigcup_{\emptyset \subset X \subseteq N_i} T_{i,X}$ and $C'' = C' \cup \bigcup_{i=1}^{q} T_i$. Observe that $C''$ contains $|T_i| \leq k|N_i|^k$ vertices of $K_i$, which implies in particular that $\mathrm{tw}(\mathrm{torso}(G[K_i], C'' \cap K_i)) \leq k\binom{|N_i|}{k}$. Therefore, by Lemma 2.9 the treewidth of $\mathrm{torso}(G, C'')$ is larger than the treewidth of $\mathrm{torso}(G, C')$ by a constant depending only on $k$, which means that it can be still bounded by a constant depending only on $k$.

Consider a connected $s-t$ separator $S$ of size at most $k$ such that $|S \setminus C''|$ is minimum possible. If $S \subseteq C''$, then there is nothing to prove so we assume that $S \setminus C''$ is non-empty. By the definition of $C'$, the set $S \cap C'$ is an $s-t$ separator as well: otherwise, $S$ contains a minimal $s-t$ separator that has a vertex outside $C'$. Consider a vertex $v \in S \setminus C''$ and suppose that $v \in K_i$. Let $T$ be the connected component of $G[S \cap (K_i \cup N_i)]$ containing $v$ and let $X = T \cap N_i$. Thus $T$ is a connected set of size at most $k$ with $X \subseteq T \subseteq K_i \cup N_i$, which means that $T_{i,X}$ is nonempty and $|T_{i,X}| \leq |T|$. Let $S' := (S \setminus T) \cup T_{i,X}$, it is clear that $|S'| \leq |S|$ and $S'$ is also an $s-t$ separator (as it contains $S \cap C'$). To see that $G[S']$ is connected, observe that $G[T_{i,X}]$ is connected and contains every vertex of $X$ by definition. Furthermore, every component of $G[S' \setminus K_i]$ contains at least one vertex of $X$, hence every vertex of $G[S']$ is in the same component. As $|S' \setminus C''| < |S \setminus C''|$, this contradicts the minimality of $S$. □

With Lemma 3.6 at hand, it is fairly simple to find a connected $s-t$ separator of size at most $k$:

**Theorem 3.7.** *Finding a connected $s-t$ separator of size at most $k$ is linear-time FPT.*

*Proof.* Let us construct the set $C''$ using the algorithm of Lemma 3.6, and let $G^* = \text{torso}(G, C'' \cup \{s,t\})$. Let us mark the edges of $G^*$ by two colors: let an edge $xy$ be "black" if it appears also in $G$, and let $xy$ be "red" if it appears only in $G^*$. According to Lemma 3.6, it remains to find an $s-t$ separator $S$ of size at most $k$ in $G^*$ such that the black edges in $G^*[S]$ form a connected subgraph that span every vertex of $S$. It is a routine exercise to formulate in monadic second order logic the problem of finding such a separator. As the treewidth of $G^*$ is bounded by a function of $k$, the linear-time algorithm follows by Courcelle's Theorem. □

## 3.4 Multicut problems

In Sections 3.1–3.3, we proved tractability results for problems where the task is to separate two vertices $s$ and $t$ with a separator of size at most $k$ that satisfies some additional constraint (e.g., induces a certain graph). In this section, we demonstrate the power of our methodology by showing that it is possible to solve problems where the underlying separation task is more complicated than separating a single pair $(s,t)$ of vertices. We generalize the problem in three ways: (1) instead of a single pair, we need to separate multiple pairs; (2) instead of pairs of vertices, we need to separate pairs of subsets of vertices; and (3) for some of the pairs, we modify the problem by requiring that the removal of the cutset *does not* separate them.

MULTICUT is the generalization of MINCUT where, instead of $s$ and $t$, the input contains a set $(s_1, t_1)$, ..., $(s_\ell, t_\ell)$ of terminal pairs. The task is to find a set $S$ of at most $k$ vertices that separates $s_i$ and $t_i$ for every $1 \leq i \leq \ell$. MULTICUT is known to be FPT [30, 50, 65] parameterized by $k$ and $\ell$. Very recently, it has been shown that the problem is FPT parameterized by $k$ only [7, 54]. (Note that all our results in this section are parameterized by both $k$ and $\ell$, thus they do not generalize the FPT results of [7, 54] parameterized by $k$ only.) In the MULTIWAY CUT problem, the input contains a set $T$ of terminals and the task is to find a set $S$ of at most $k$ vertices that pairwise separate the vertices in $T$. Let us observe that MULTIWAY CUT is a special case of MULTICUT. This special case has been proved to be FPT parameterized by $k$ only already in [11, 50].

We define SET-MULTICUT to be the variant where instead the pairs $(s_1, t_1)$, ..., $(s_\ell, t_\ell)$, we are given pairs $(X_1, Y_1), \ldots, (X_\ell, Y_\ell)$ of sets and want to find a set $S$ of at most $k$ nonterminal vertices that separate $X_i$ and $Y_i$ for every $1 \leq i \leq \ell$. Note that in this variant the set $S$ is allowed to intersect $X_i$ or $Y_i$. In many cases, it is easy to reduce the problem of separating two sets to the problem of separating two vertices: if we add a new vertex $s_i$ adjacent to $X_i$ and a new vertex $t_i$ adjacent to $Y_i$, then a set disjoint from $\{s_i, t_i\}$ is an $s_i - t_i$ separator if and only if it separates $X_i$ and $Y_i$ in the original graph. This might suggest that the fixed-parameter tractability of SET-MULTICUT problem parameterized by $k$ and $\ell$ easily follows from this kind of fixed-parameter tractability for the ordinary multicut problem: for every $1 \leq i \leq \ell$, add a vertex $s_i$ adjacent to each $X_i$ and a vertex $t_i$ adjacent to each $Y_i$, consider $(s_1,t_1),\ldots(s_\ell,t_\ell)$ as pairs of terminals and solve the resulting instance of the multicut problem (with the additional constraint that the solution is disjoint from $\{s_1,\ldots,s_\ell,t_1,\ldots,t_\ell\}$). However this approach, which works very well for $\ell = 1$ fails even for $\ell = 2$, see Figure 2. Thus the parameterization of the SET-MULTICUT problem by $k$ and $\ell$ is *not* a trivial extension of the parameterization by $k$ and $\ell$ of the ordinary multicut problem.

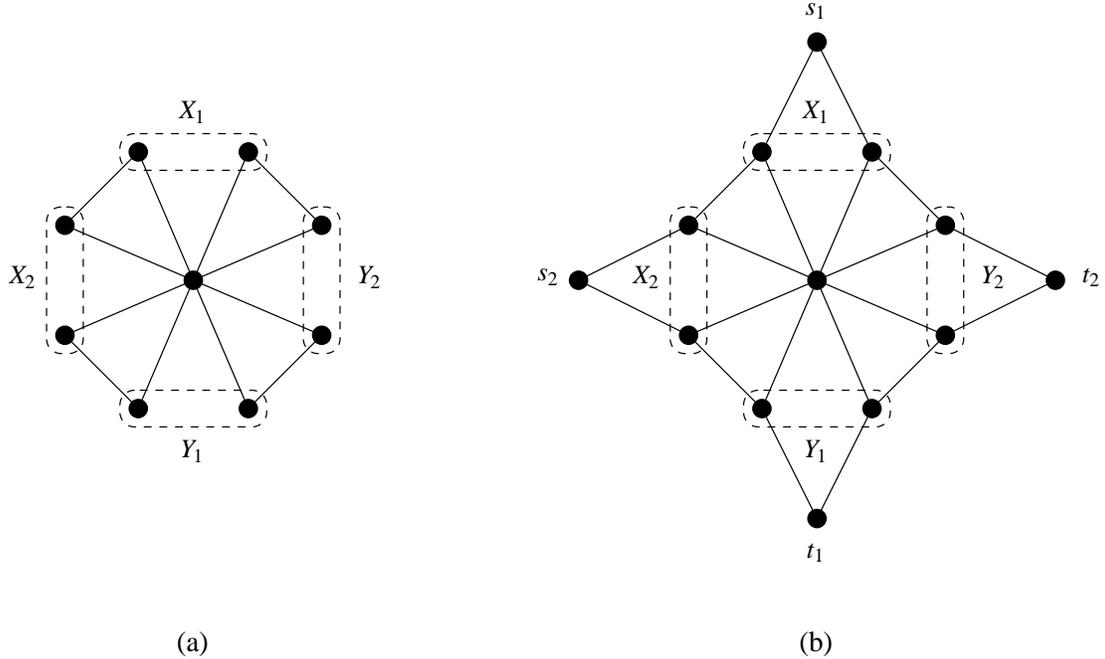

Figure 2: (a) Removing the central vertex separates $X_1$ from $Y_1$ and $X_2$ from $Y_2$. (b) Removing the central vertex *does not* separate $s_1$ from $t_1$ and $s_2$ from $t_2$.

We further generalize SET-MULTICUT by defining SET-MULTICUT-UNCUT to be the variant where the input contains an additional integer $\ell' \leq \ell$, and we change the problem by requiring for every $\ell' \leq i \leq \ell$ that $S$ *does not* separate $X_i$ and $Y_i$. Finally, if $\mathcal{G}$ is a class of graphs, then (analogously to Section 3.1) we define $\mathcal{G}$-SET-MULTICUT-UNCUT to be the problem with the additional requirement that the solutions $S$ induces a member of $\mathcal{G}$. The main result of the section is the following:

**Theorem 3.8.** *If $\mathcal{G}$ is* decidable *and* hereditary, *then $\mathcal{G}$-MULTICUT-UNCUT is linear-time FPT parameterized by $k$ and $\ell$.*

For the proof of Theorem 3.8, we need to adapt some statements of Section 2 to the setting of separating sets of vertices. First, we restate Lemma 2.11 in terms of separating two sets $X$ and $Y$.

**Lemma 3.9.** *Let $X,Y$ be sets of vertices of $G$. For some $k \geq 0$, let $C$ be the union of all minimal sets $S$ of size at most $k$ separating $X$ and $Y$. There is a $O(f(k) \cdot (|E(G)|+V(G)))$ time algorithm that returns a set $C' \supseteq C$ such that the treewidth of $\mathrm{torso}(G,C')$ is at most $g(k)$, for some functions $f$ and $g$ of $k$.*

*Proof.* Let $G'$ be the graph obtained by introducing a new vertex $s$ adjacent to $X$ and a new vertex $t$ adjacent to $Y$. It is clear that a set $S \subseteq V(G)$ is an $s-t$ separator in $G'$ if and only if it separates $X$ and $Y$ in $G$. Thus Lemma 2.11 gives us a set $C' \subseteq V(G')$ disjoint from $\{s,t\}$ that fully contains every minimal set of size at most $k$ separating $X$ and $Y$ in $G$. As $G$ is a subgraph of $G'$, the treewidth of $\mathrm{torso}(G,C')$ is at most the treewidth of $\mathrm{torso}(G',C')$. □

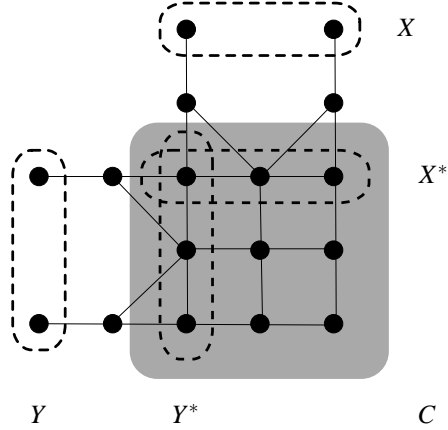

Figure 3: Schematic illustration of the sets introduced in Lemma 3.10

Next we adapt Prop. 2.7 to the case of separating sets of vertices. The only technical detail is to adjust the sets $X$ and $Y$ in order to account for those vertices that are not in the torso.

**Lemma 3.10.** *Let $G$ be a graph and $X, Y, C \subseteq V(G)$ sets of vertices such $C$ separates $X$ and $Y$. Let us define $X^*$ such that $v \in C$ is in $X^*$ if and only if there is a path $P$ between $v$ and a vertex $w \in X$ such that $v$ is the only vertex of $P$ in $C$ (in particular, $X \cap C \subseteq X^*$). Let $Y^*$ be defined analogously. (For intuitive illustration of these sets, see Figure 3.) Then for every $S \subseteq C$, the set $S$ separates $X$ and $Y$ in $G$ if and only it separates $X^*$ and $Y^*$ in $\mathrm{torso}(G, C)$.*

*Proof.* Consider a set $S \subseteq C$ that *does not* separate $X^*$ and $Y^*$ in $\mathrm{torso}(G, C)$: by Prop 2.7, this implies that there is a path $P$ between some $x \in X^*$ and $y \in Y^*$ in $G \setminus S$. Let $P_x$ be the path that connects $x$ and vertex of $X$ as in the definition of $X^*$, and let $P_y$ be the similar path for $Y$; note that $P_x$ and $P_y$ are disjoint from $S \subseteq C$. Now the path $P_x P P_y$ connects a vertex of $X$ and a vertex of $Y$ in $G \setminus S$, i.e., $S$ does not separate $X$ and $Y$ in $G$.

Consider now a set $S \subseteq C$ that *does not* separate some $x \in X$ and $y \in Y$ in $G$, and let $P$ a path connecting $x$ and $y$ in $G \setminus S$. As $C$ separates $X$ and $Y$, the path $P$ intersects $C$. Going from $x$ to $y$ on $P$, the first vertex of $C$ is in $X^*$ (denote it by $x^*$) and the last vertex of $C$ is in $Y^*$ (denote it by $y^*$). As $S$ does not separate $x^*$ and $y^*$ in $G$, Prop. 2.7 implies that $S$ does not separate $x^*$ and $y^*$ (and hence $X^*$ and $Y^*$) in $\mathrm{torso}(G, C)$. □

Now we are ready to prove the fixed-parameter tractability of $\mathcal{G}$-SET-MULTICUT-UNCUT

*Proof of Theorem 3.8.* Consider an instance $(G, \{(X_1, X_2), \ldots, (X_\ell, Y_\ell)\}, k, \ell')$ of $\mathcal{G}$-SET-MULTICUT-UNCUT. For every $1 \leq i \leq \ell'$, let us use Lemma 3.9, to obtain a set $C'_i$ that contains every set of size at most $k$ that separates $X_i$ and $Y_i$. Note that we can assume that $C'_i$ separates $X_i$ and $Y_i$: otherwise, there is no set of size at most $k$ that separates $X_i$ and $Y_i$, and we can return "NO." Let $C = \bigcup_{i=1}^{\ell'} C'_i$; by Lemma 2.8, the treewidth of $\mathrm{torso}(G, C)$ can be bounded by a constant depending only on $k$ and $\ell'$. Any minimal solution of $\mathcal{G}$-SET-MULTICUT-UNCUT having size at most $k$ is a subset of $C$: if a vertex of the solution is not part of a minimal set separating $X_i$ and $Y_i$ for some $1 \leq i \leq \ell'$, then removing this vertex from the solution cannot make the solution invalid; in particular, no pair $(X_i, Y_i)$ with $\ell' < i \leq \ell$ can become separated after the removal. Therefore, if $C$ does not separate $X_i$ and $Y_i$ for some $\ell' < i \leq \ell$, then we can remove the pair $(X_i, Y_i)$ from consideration, as it is not separated by any minimal solution. Thus in the following, we can assume that $C$ separates $X_i$ and $Y_i$ for every $1 \le i \le \ell$.

For $1 \le i \le \ell$, let us define $X_i^*$ and $Y_i^*$ as in Lemma 3.10. By Lemma 3.10, a set $S \subseteq C$ separates $X_i$ and $Y_i$ in $G$ if and only $C$ separates $X_i^*$ and $Y_i^*$ in torso$(G,C)$. Let us mark the edges of torso$(G,C)$ by two colors: let an edge $xy$ be "black" if it appears also in $G$, and let $xy$ be "red" if it appears only in torso$(G,C)$. Now our task is to find a set $S \subseteq C$ of size at most $k$ such that the black edges of $G[S]$ for a member of $\mathcal{G}$, the set $S$ separates $X_i^*$ and $Y_i^*$ in torso$(G,C)$ for every $1 \le i \le \ell'$, and $S$ *does not* separate $X_i^*$ and $Y_i^*$ in torso$(G,C)$ for $\ell' < i \le \ell$. It is a routine exercise to formulate in monadic second order logic the problem of finding such a set $S$; only some straightforward modifications are required compared to the formulations in Theorems 3.1 and 3.7 (e.g., we need to introduce a unary predicate describing each set $X_i^*$, $Y_i^*$). As the treewidth of torso$(G,C)$ is bounded by a function of $k$ and $\ell'$, the linear-time algorithm follows by Courcelle's Theorem. Notice that the treewidth of *torso$(G,C)$* does not depend on $\ell$, i.e., by the number of pairs that should not be separated. The dependence of the runtime on $\ell$ rather than on $\ell'$ is caused by the size of the resulting monadic second order logic formula. □

Theorem 3.8 helps clarifying a theoretical issue. In Section 2, we defined $C$ as the set of all vertices appearing in minimal $s-t$ separators of size at most $k$. There is no obvious way of finding this set in FPT-time and Lemma 2.8 produces only a superset $C'$ of $C$. However, Theorem 3.8 can be used to find $C$: a vertex $v$ is in $C$ if and only if there is a set $S$ of size at most $k-1$ and two neighbors $v_1, v_2$ of $v$ such that $S$ separates $s$ and $t$ in $G \setminus v$, but $S$ does not separate $s$ from $v_1$ and $t$ from $v_2$ in $G \setminus v$ (including the possibility that $v_1 = s$ or $v_2 = t$).

## 4 Constrained Bipartization Problems

Reed et al. [60] solved a longstanding open question by proving the fixed-parameter tractability of the BIPARTIZATION problem: given a graph $G$ and an integer $k$, find a set $S$ of at most $k$ vertices such that $G \setminus S$ is bipartite (see also [49] for a somewhat simpler presentation of the algorithm). This paper introduced the technique of "iterative compression," which has become a standard tool in parameterized complexity [63]. The running time of this algorithm is $O(k \cdot 3^k |V(G)| \cdot |E(G)|)$, i.e., quadratic time for fixed $k$. Very recently, a significantly more complex new algorithm with almost linear running time were presented by Kawarabayashi and Reed [44] (throughout this paper, *almost linear* means that the running time is $O(|E(G)|\alpha(|E(G)|, |V(G)|))$, where $\alpha$ is the inverse of the Ackermann function).

In this section we consider the $\mathcal{G}$-BIPARTIZATION problem: a generalization of BIPARTIZATION where, in addition to $G \setminus S$ being bipartite, it is also required that $S$ induces a graph belonging to a class $\mathcal{G}$. We prove that given *any* solution $S_0$ of size at most $k$ for BIPARTIZATION, our algorithm for $\mathcal{G}$-MINCUT (Section 3.1) can be used to check in linear time (for fixed $k$) whether or not there exists a solution $S$ of size at most $k$ for BIPARTIZATION such that $G[S]$ belongs to $\mathcal{G}$. Thus the quadratic-time and almost linear-time algorithms of [60] and [44] imply quadratic-time and almost linear-time algorithms, respectively, for $\mathcal{G}$-BIPARTIZATION.

A key idea in the algorithm of Reed et al. [60] is that if a set $X$ is given such that $G \setminus X$ is bipartite, then BIPARTIZATION can be solved by at most $3^{|X|}$ applications of a procedure solving MINCUT. The following lemma allows us to transform BIPARTIZATION into a separation problem.

**Lemma 4.1.** *Let $G$ be a bipartite graph and let $(B', W')$ be a 2-coloring of the vertices. Let $B$ and $W$ be two subsets of $V(G)$. Then for any set $S \subseteq V(G)$, the graph $G \setminus S$ has a 2-coloring where $B \setminus S$ is black and $W \setminus S$ is white if and only if $S$ separates $X := (B \cap B') \cup (W \cap W')$ and $Y := (B \cap W') \cup (W \cap B')$.*

*Proof.* In a 2-coloring of $G \setminus S$, each vertex either has the same color as in $(B', W')$ (call it an unchanged vertex) or the opposite color as in $(B', W')$ (call it a changed vertex). Observe that a changed and an unchanged vertex cannot be adjacent: in this case, they would have the same color either under $(B', W')$ or under the considered coloring of $G \setminus S$. Consequently, a changed and an unchanged vertex cannot belong to the same connected component of $G \setminus S$, because this would imply existence of an edge between a changed and an unchanged vertex. If $B$ is black and $W$ is white in a 2-coloring of $G \setminus S$, then clearly $X \setminus S$ is unchanged and $Y \setminus S$ is changed. Thus $S$ has to separate $X$ and $Y$ in $G$.

For the other direction, suppose that $X \setminus S$ is separated from $Y \setminus S$ in $G \setminus S$. We modify the coloring $(B', W')$ by changing the color of every vertex that is in the same connected component of $G \setminus S$ as some vertex of $Y$. Since the vertices of the same component either all change their colors or all remain colored in the same color as in $(B', W')$, the resulting coloring is a proper 2-coloring of $G \setminus S$. By construction, all vertices of $Y$ have the desired color. Since $S$ separates $X$ and $Y$, the vertices of $X \setminus S$ are unchanged and hence have the required colors as well. □

The main result of the section is the following:

**Theorem 4.2.** *If $\mathcal{G}$ is hereditary and decidable, then $\mathcal{G}$-BIPARTIZATION is almost linear-time FPT parameterized by $k$.*

*Proof.* Using the algorithm of [44], we first try to find a set $S_0$ of size at most $k$ such that $G \setminus S_0$ is bipartite. If no such set exists, then clearly there is no set $S$ satisfying the requirements. Otherwise, we branch into $3^{|S_0|}$ directions: if we fix a hypothetical solution $S$ and a 2-coloring of $G \setminus S$, then each vertex of $S_0$ is either removed (i.e., in $S$), colored black, or colored white. For a particular branch, let $R = \{v_1, \ldots, v_r\}$ be the vertices of $S_0$ to be removed and let $B_0$ (resp., $W_0$) be the vertices of $S_0$ having color black (resp., white) in a 2-coloring of the resulting bipartite graph. We say that a set $S$ is *compatible* with partition $(R, B_0, W_0)$ if $S \cap S_0 = R$ and $G \setminus S$ has a 2-coloring where $B_0$ and $W_0$ are colored black and white, respectively. It is easy to see that $(G, k)$ is a yes-instance of the $\mathcal{G}$-BIPARTIZATION problem if and only if for at least one branch corresponding to partition $(R, B_0, W_0)$ of $S_0$, there is a set $S$ compatible with $(R, B_0, W_0)$ having size at most $k$ and such that $G[S] \in \mathcal{G}$. Clearly, we need to check only those branches where $G[B_0]$ and $G[W_0]$ are both independent sets.

We transform finding a set compatible with $(R, B_0, W_0)$ into a separation problem. Let $(B', W')$ be a 2-coloring of $G \setminus S_0$. Let $B = N(W_0) \setminus S_0$ and $W = N(B_0) \setminus S_0$. Let us define $X$ and $Y$ as in Lemma 4.1, i.e., $X := (B \cap B') \cup (W \cap W')$, and $Y := (B \cap W') \cup (W \cap B')$. We construct a graph $G'$ that is obtained from $G$ by deleting the set $B_0 \cup W_0$, adding a new vertex $s$ adjacent with $X \cup R$, and adding a new vertex $t$ adjacent with $Y \cup R$. Note that every $s-t$ separator in $G'$ contains $R$. By Lemma 4.1, a set $S$ is compatible with $(R, B_0, W_0)$ if and only if $S$ is an $s-t$ separator in $G'$. Thus what we have to decide is whether there is an $s-t$ separator $S$ of size at most $k$ such that $G'[S] = G[S]$ is in $\mathcal{G}$. That is, we have to solve the $\mathcal{G}$-MINCUT instance $(G', s, t, k)$. The fixed-parameter tractability of the $\mathcal{G}$-BIPARTIZATION problem now immediately follows from Theorem 3.1. □

**Remark 4.3.** Instead of the very complex almost-linear time algorithm of [44], one can use the much simpler quadratic-time algorithm of [60] to find the set $S_0$. This increases the total running time from almost linear to quadratic. There is a similar possible trade off between running time and simplicity in all the results in remaining part of the current paper which rely on [44].

Similar to Theorem 3.1, we can generalize Theorem 4.2 such that $\mathcal{G}$ contains graphs colored with a fixed finite number of colors.

Theorem 4.2 immediately implies that the STABLE BIPARTIZATION problem ("Is there an independent set of size at most $k$ whose removal makes the graph bipartite?") is FPT: just set $\mathcal{G}$ to be the class of all graphs without edges. This answers an open question of Fernau [14]. Next, we show that the EXACT STABLE BIPARTIZATION problem ("Is there an independent set of size *exactly* $k$ whose removal makes the graph bipartite?") is also FPT, answering a question posed by Díaz et al. [16]. An obvious approach would be to modify the algorithm of Theorem 4.2 such that we find an independent $s-t$ separator $S$ of size exactly $k$ (instead of size at most $k$). However, finding such a separator is W[1]-hard by Theorem 3.3, making this approach unlikely to work. Instead, we argue that under appropriate conditions, any solution of size at most $k$ can be extended to an independent set of size exactly $k$. We find it somewhat surprising that for independent $s-t$ separators finding a solution of size exactly $k$ is harder than the finding a solution of size at most $k$, while in the closely related bipartization problem the two variants have the same complexity.

For the proof, we use the folklore result that a minimum length odd cycle in a graph $G$ can be found in polynomial time. We need this statement in the following form:

**Lemma 4.4.** *Given a graph $G$ and a set $S \subseteq V(G)$ such that $G \setminus S$ is bipartite, we can find a minimum length odd cycle in time $O(|S|(|E(G)|+|V(G)|))$.*

*Proof.* Let us construct a bipartite graph $G'$ where two vertices $v_1$, $v_2$ correspond to each vertex $v$ of $G$ and two edges $v_1 u_2$ and $v_2 u_1$ correspond to each edge $uv$ of $G$. We claim that $G$ has an odd cycle of length at most $k$ if and only for some $v \in S$ there is a path of length at most $k$ between $v_1$ and $v_2$ in $G$. Testing the later condition can be done by performing a breadth-first search starting from $v_1$ for every $v \in S$, which takes linear-time per each vertex of $S$.

To prove the claim, let us observe first that graph $G'$ is bipartite, and if $v_1$ and $v_2$ are in the same component of $G'$ then every path between $v_1$ and $v_2$ has odd length. If $G$ has an odd cycle of length at most $k$, then it has to go through some $v \in S$ and it is easy to see that there is a corresponding path of length $k$ from $v_1$ to $v_2$ in $G'$. Conversely, given a path $P$ of (odd) length $k$ from $v_1$ to $v_2$ in $G'$, there is a corresponding closed walk $P'$ of (odd) length $k$ in $G$ and this implies that a subwalk of $P'$ is an odd cycle of length at most $k$ in $G$. □

**Theorem 4.5.** *Given a graph $G$ and an integer $k$, deciding whether $G$ can be made bipartite by the deletion of an independent set of size exactly $k$ is almost linear-time FPT.*

*Proof.* It is more convenient to consider an annotated version of the problem where the independent set being deleted is required to be a subset of a set $D \subseteq V(G)$ given as part of the input. To express the original problem without the annotation, $D$ is initially set to $V(G)$. The algorithm has the following 4 stopping conditions.

- If $k = 0$ and $G$ is bipartite, then return "YES."

- If $k = 0$, but $G$ is not bipartite, then return "NO."

- If $k > 0$, but $G$ is bipartite, then decide in a polynomial time whether $G[D]$ has an independent set of size exactly $k$.

- If $k > 0$ and $G \setminus D$ is not bipartite, then return "NO."

Assume that none of the above conditions is satisfied. Then the algorithm starts by finding an odd cycle $C$ of minimum length. For this purpose, we first invoke the algorithm of Kawarabayashi and Reed [44] to find a set $S$ of at most $k$ vertices such that $G \setminus S$ is bipartite (note that we do not require here that $S$ is in $D$). If there is no such set $S$, then we can return "NO." Otherwise, we can use Lemma 4.4 to find a shortest odd cycle $C$.

It is not difficult to see that the minimality of $C$ implies that either $C$ is a triangle or $C$ is chordless. Moreover, in the latter case, every vertex $v$ not in $C$ is adjacent to at most 2 vertices of the cycle. To see this, note first that if the length of $C$ is more than 3, then the minimality of $C$ implies that $v$ cannot be adjacent with two adjacent vertices of $C$ (as they would form a triangle). Thus if $v$ has at least 3 (nonadjacent) neighbors in $C$, then the length of $C$ is at least 7 and $v$ has two neighbors $x$ and $y$ whose distance in $C$ is at least 3. Vertices $x$ and $y$ split $C$ into a path of odd length and a path of even length. Replacing the even-length path (whose length is at least 4) with the path $xvy$ of length 2 gives a shorter odd cycle, contradicting the minimality of $C$.

Since none of the stopping conditions holds, $|V(C) \cap D| > 0$. If $1 \leq |V(C) \cap D| \leq 3k+1$, then we branch on the selection of each vertex $v \in V(C) \cap D$ into the set $S$ of vertices being removed and apply the algorithm recursively with the parameter $k$ being decreased by 1 and the set $D$ being updated by removal of $v$ and $N(v) \cap D$. If $|V(C) \cap D| > 3k+1$, then we apply the approach of Theorem 4.2 to find an independent set $S$ of size at most $k$ whose removal makes the graph bipartite. To ensure that $S \subseteq D$ we may, for example split all vertices $v \in V(G) \setminus D$ into $k+1$ independent copies with the same neighborhood as $v$. If $|S| = k$, we are done. Otherwise, $|S| = k' < k$. In this case we observe that by construction each vertex of $S$ (either in $C$ or outside $C$) forbids the selection of at most 3 vertices of $V(C) \cap D$ (including itself, if it is in $C$). Thus the number of vertices of $V(C) \cap D$ allowed for selection is at least $3k+1-3k' = 3(k-k')+1$. Since the cycle is chordless, we can select $k-k'$ independent vertices among them and thus complement $S$ to being of size exactly $k$. Therefore, if the algorithm succeeds to find an independent set $S$ of size at most $k$ whose removal makes the graph bipartite, it may safely return "YES." It is clear that otherwise "NO" can be returned. □

## 4.1 Edge bipartization

It is equally natural to study the edge-deletion version of bipartization ("Given graph $G$ and integer $k$, can $G$ be made bipartite by the deletion of at most $k$ edges?"). Analogously to $\mathcal{G}$-BIPARTIZATION, we introduce a constrained version of the problem, where the edges removed need to form a graph belonging to a certain class. Formally, if $\mathcal{G}$ is a class of graphs, $\mathcal{G}$-EDGEBIPARTIZATION problem asks if $G$ has a subgraph $H$ such that $H$ has at most $k$ edges, $H \in \mathcal{G}$, and $G \setminus E(H)$ (i.e., removing the edges of $H$) is bipartite. By a direct reduction to the vertex-deletion variant, we show that this problem is FPT whenever $\mathcal{G}$ is decidable and closed under taking subgraphs. In Section 4.2, the fixed-parameter tractability of BIPARTITE CONTRACTION is obtained as an easy corollary of this result.

The reduction is conceptually simple, but somewhat technical to describe. It is convenient to use the generalization of Theorem 4.2 which allows graphs to have a finite number of labels (colors) on the vertices. To describe the reduction, we need the following transformation. For every graph $G$, we define a graph $G'$ where the vertices have labels from the set $\{1,2,3\}$:

- For every $v \in V(G)$, let $V(G')$ contain a label-1 vertex $v^1$, a label-2 vertex $v^2$, let $E(G')$ contain the edge $v^1 v^2$,

- For every $uv \in E(G)$, let $V(G')$ contain two label-3 vertices $e', e''$, one of them adjacent to $u^1$ and $v^2$, the other one adjacent to $u^2$ and $v^1$.

The following statement is easy to see (note that the induced subgraph relation considered here respects the labels):

**Proposition 4.6.** *For two undirected graphs $G$ and $H$, $G'$ is an induced subgraph of $H'$ if and only if $G$ is a subgraph of $H$.*

*Proof.* The "if" direction is obvious: the edges/vertices that are present in $H$ but not in $G$ correspond to vertices present in $H'$ but missing from $G'$. For the "only if" direction, observe that every label-1 vertex in $G'$ or $H'$ has a unique label-2 neighbor. Therefore, for every $v \in V(G)$ every induced subgraph embedding of $G'$ into $H'$ maps $v^1, v^2 \in V(G')$ to $w^1, w^2 \in V(H')$ for some $w \in V(H)$. Using this mapping, it is easy to see that every vertex or edge of $G$ has a corresponding vertex or edge in $H$, i.e., $G$ is a subgraph of $H$. □

**Theorem 4.7.** *If $\mathcal{G}$ is decidable and closed under taking subgraphs, then $\mathcal{G}$-EDGEBIPARTIZATION is almost linear-time FPT parameterized by $k$.*

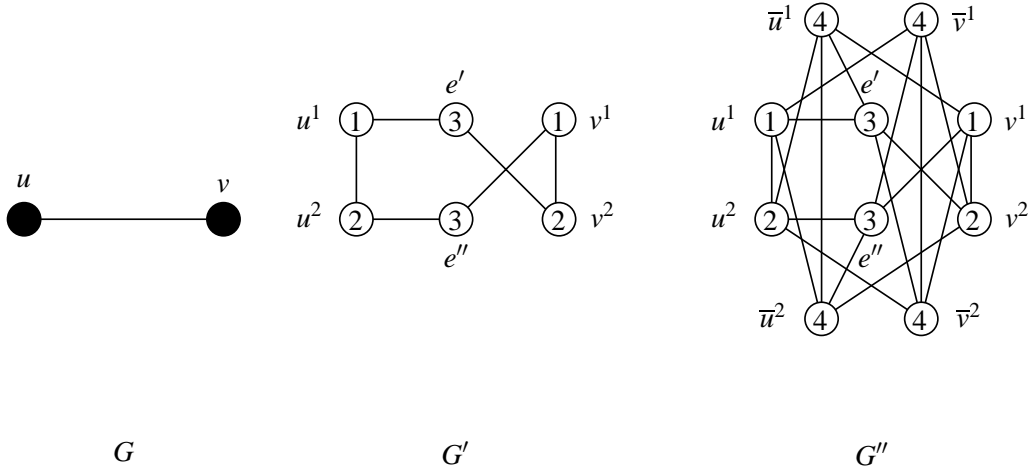

Figure 4: Constructing the graph $G'$ and $G''$ into proof Theorem 4.7.

*Proof.* Given an instance $(G,k)$ of $\mathcal{G}$-EDGEBIPARTIZATION, we reduce it to an instance of $\mathcal{G}_k$-BIPARTIZATION for an appropriate (finite) class $\mathcal{G}_k$, constructed as follows. For every graph $H \in \mathcal{G}$ having at most $k$ edges, let $\mathcal{G}_k$ contain $H'$ and every induced subgraph of $H'$. By definition, $\mathcal{G}_k$ is hereditary. Note that if $H \notin \mathcal{G}$, then Prop. 4.6 implies that $H' \notin \mathcal{G}_k$.

Let us construct $G'$ from $G$. We extend $G'$ to obtain a graph $G''$ the following way. For every $v \in V(G)$, we introduce two new adjacent vertices $\bar{v}^1$, $\bar{v}^2$ having label 4. For $i = 1,2$, we make $\bar{v}^i$ adjacent to every neighbor of $v^i$ in $G'$. Furthermore, for every neighbor $u$ of $v$ in $G$, we make $\bar{v}^i$ adjacent to $u^i$.

We claim that $(G,k)$ is a yes-instance of $\mathcal{G}$-EDGEBIPARTIZATION if and only if $G''$ has a set $S$ such that $G'' \setminus S$ such that $G''[S] \in \mathcal{G}_k$. Suppose first that $G$ has a subgraph $H$ with at most $k$ edges such that $G \setminus E(H)$ is bipartite and $H$ is in $\mathcal{G}$. Then $H' \in \mathcal{G}_k$ by definition of $\mathcal{G}_k$. For every $v \in V(H)$, let $S$ contain the vertices $v^1$ and $v^2$; for every $e \in E(H)$, let $S$ contain the label-3 vertices $e', e''$ corresponding to $e$. Observe that $G''[S]$ is isomorphic to a supergraph of $H' \in \mathcal{G}_k$: the additional edges are due to edges of $G$ that are not in $E(H)$, but connect two vertices of $V(H)$. Therefore, $G''[S]$ is in $\mathcal{G}_k$.

We show that $G'' \setminus S$ is bipartite. Consider a 2-coloring of $G \setminus E(H)$. If $v \in V(G)$ has color $i$, then let $v^1$, $\bar{v}^1$ have color $i$ and $v^2$, $\bar{v}^2$ have color $3-i$ (if they appear in $G'' \setminus S$). Observe that this is a proper 2-coloring of these vertices: for example, if $\bar{u}^1$ and $v^1$ are adjacent in $G'' \setminus S$ and have the same color, then $u$ and $v$ are adjacent in $G$ and have the same color, which means that $uv \in E(H)$, and therefore $v^1 \in S$, a contradiction. Furthermore, for every label-3 vertex in $G'' \setminus S$, the neighbors have the same color, thus the coloring can be extended to the label-3 vertices. To see this, recall that if a vertex $e'$ of $G'' \setminus S$ represents an edge $e = uv \in E(G)$, then its neighborhood is in $\{u^i, \bar{u}^i, v^{3-i}, \bar{v}^{3-i}\}$ for some $i = 1,2$. The fact that $e'$ is in $G'' \setminus S$ implies that $uv \notin E(H)$, and therefore $u$ and $v$ have different colors in the 2-coloring of $G \setminus E(H)$. This means that $\{u^i, \bar{u}^i, v^{3-i}, \bar{v}^{3-i}\}$ indeed have the same color.

Suppose now that $G'' \setminus S$ is bipartite and $G''[S]$ is in $\mathcal{G}_k$. As $S$ contains no vertices having label 4, we can define a (not proper) 2-color of $G$ by assigning the color of $\bar{v}^1$ to $v$. Let $H$ be the subgraph of $G$ spanned by those edges whose endpoints have the same color. Clearly, $G \setminus E(H)$ is bipartite. For every edge $e = uv \in E(H)$, the two corresponding label-4 vertices $e', e''$ in $G''$ are in $S$: these vertices are adjacent to $\bar{u}^i$ and $\bar{v}^{3-i}$ (for some $i = 1,2$) whose colors are different. Similarly, for every $uv \in E(H)$, the vertices $u^1$, $u^2$, $v^1$, $v^2$ have to be in $S$, as each of them are adjacent to vertices of both colors. Therefore, $G''[S] \in \mathcal{G}_k$ contains $H'$ as an induced subgraph. Since $\mathcal{G}_k$ is hereditary, it follows that $H' \in \mathcal{G}_k$. As we have observed at the beginning of the proof, this is only possible if $H \in \mathcal{G}$. Thus $(G,k)$ is indeed a yes-instance of $\mathcal{G}$-BIPARTIZATION. □

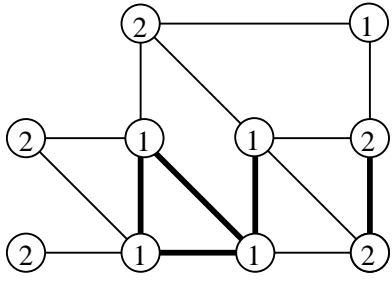

Figure 5: A graph that can be made bipartite by removing the 5 strong edges (the numbers on the vertices show the resulting 2-coloring). The strong edges span a graph with rank 4. Therefore, the graph can be made bipartite by contracting four edges: contract each of the two components into a single vertex.

## 4.2 Bipartite contraction

If $e$ is an edge of graph $G$, then the *contraction* of $e$ in $G$ is the graph $G/e$ obtained by identifying the endpoints of $e$ and removing loops and parallel edges. If $S$ is a set of edges in $G$, then we denote by $G/S$ the graph obtained by contracting the edges in $S$ in an arbitrary order until all of them are removed.

Problems defined by deleting the minimum number of edges or vertices to achieve a certain property have been intensively studied in the literature, especially from the viewpoint of fixed-parameter tractability [9,31,41,44–47,51,60,66]. Recently, there has been increased interest in analogous problems defined by *contractions* [2,3,27,36–38,42,64]. In many respects, deletions and contractions behave very differently. For example, Heggernes et al. [38] showed that BIPARTITE CONTRACTION (Given a graph $G$ and integer $k$, can $G$ be made bipartite by the contraction of at most $k$ edges?) is FPT, but their algorithm is significantly more complex than the relatively simple algorithms for the bipartite edge/vertex deletion problems [31, 49, 60]. On the other hand, it seems that contraction problems fit naturally into our study of generalized bipartization. Using a simple combinatorial observation, we obtain the fixed-parameter tractability of BIPARTITE CONTRACTION as a corollary.

Let us denote by $r(G)$ the *rank* of graph $G$, i.e., the number of edges in a spanning forest of $G$ (equivalently, the number of vertices minus the number of connected components). Note that $r(G') \le r(G)$ for every subgraph $G'$ of $G$. The following lemma allows us to translate bipartite contraction into an edge deletion problem (see Figure 5):

**Lemma 4.8.** *For every graph $G$ and integer $k$, the following statements are equivalent:*

(1) *$G$ has a subgraph $F$ such that $|E(F)| \le k$ and $G/E(F)$ is bipartite, and*
(2) *$G$ has a subgraph $H$ such that $r(H) \le k$ and $G \setminus E(H)$ is bipartite.*

*Proof.* (1) $\Rightarrow$ (2): Let us extend $F$ to a subgraph $H$ of $G$ that includes every edge of $G$ whose endpoints are in the same connected component of $F$. Clearly, $r(H) \le k$, as $F$ is a spanning forest of $H$. We claim that $G \setminus E(H)$ is bipartite. Observe that each component $K$ of $F$ is an independent set in $G \setminus E(H)$. Therefore, a 2-coloring of $G/E(F)$ can be turned into a 2-coloring of $G \setminus E(H)$ if we color every vertex in a connected component $K$ of $F$ by the color of the single vertex corresponding to $K$ in $G/E(F)$.

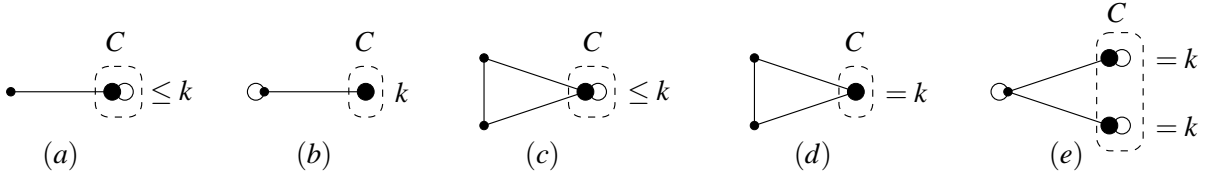

Figure 6: $(H,C,K)$- (or $(H,C,\leq K)$-) coloring with these graphs is equivalent to finding (a) a vertex cover of size at most $k$, (b) an independent set of size $k$, (c) a bipartization set of size at most $k$, (d) an independent bipartization set of size exactly $k$, (e) a bipartite independent set of size $k+k$.

(2) $\Rightarrow$ (1): Let us fix a 2-coloring of $G\setminus E(H)$. We can assume that $H$ is minimal, i.e., $G\setminus E(H')$ is not bipartite for any proper subgraph $H'$ of $H$ (here we use that $r(H)$ is monotone under taking subgraphs). Therefore, in each connect component of $H$, every vertex has the same color in the 2-coloring of $G\setminus E(H)$. This means that contracting each connected component of $H$ to a single vertex creates a bipartite graph. Such a contraction can be achieved by contracting the edges of a spanning forest $F$ of $H$, which has $r(H)\leq k$ edges. □

By Lemma 4.8, BIPARTITE CONTRACTION is equivalent to $\mathcal{G}_k$-EDGEBIPARTIZATION, where $\mathcal{G}_k$ is the set of all graphs $G$ with $r(G)\leq k$. Thus fixed-parameter tractability follows from Theorem 4.7.

**Theorem 4.9.** BIPARTITE CONTRACTION *is almost linear-time FPT.*

## 5 $(H,C,K)$-coloring

Constrained bipartization can be also considered in terms of $(H,C,K)$-*coloring*. $H$-coloring (cf. [39]) is a generalization of ordinary vertex coloring: given graphs $G$ and $H$, an $H$-*coloring* of $G$ is a homomorphism $\theta:V(G)\to V(H)$, that is, if $u,v\in V(G)$ are adjacent in $G$, then $\theta(u)$ and $\theta(v)$ are adjacent in $H$ (including the possibility that $\theta(u)=\theta(v)$ is a vertex of $H$ having a loop). It is easy to see that a graph is $k$-colorable if and only if it has a $K_k$-coloring. A seminal dichotomy result of Hell and Nešetřil [40] characterizes the complexity of finding an $H$-coloring for every $H$: it is polynomial-time solvable if $H$ is bipartite or has a loop, and NP-hard otherwise.

Various generalizations of $H$-coloring were explored in the literature, and it was possible to obtain dichotomy theorems in many cases [8, 22, 23, 26, 33–35, 39, 40]. Here we study the version of the problem allowing cardinality constraints [16–19] where, for certain vertices $v\in V(H)$, we have a restriction on how many vertices of $G$ can map to $v$. Formally, let $C\subseteq V(H)$ be a subset of vertices and let $K$ be a mapping from $C$ to $\mathbb{Z}^+$. An $(H,C,K)$-*coloring* of $G$ is an $H$-coloring with the additional restriction that $|\theta^{-1}(v)|=K(v)$ for every $v\in C$. $(H,C,\leq K)$-*coloring* is the variant of the problem where we require $|\theta^{-1}(v)|\leq K(v)$, i.e., vertex $v$ can be used *at most* $K(v)$ times. As shown in Fig. 6 and discussed in [16], this is a very a versatile problem formulation: these coloring requirements can express a wide range of fundamental problems such as $k$-INDEPENDENT SET, $k$-VERTEX COVER, BIPARTIZATION, and (EXACT) STABLE BIPARTIZATION.

Following [16], we consider the parameterized version of $(H,C,K)$-coloring, where the parameter is $k:=\sum_{v\in C}K(v)$, the total number of times the vertices with cardinality constrains can be used. Díaz et al. [16] started the program of characterizing the easy and hard cases of $(H,C,K)$- and $(H,C,\leq K)$-coloring. We make progress in this direction by showing that $(H,C,\leq K)$-coloring is FPT whenever $H\setminus C$ consists of two adjacent vertices without loops. The reader might consider this result as only a humble step towards a full dichotomy, but let us note that, as shown in Fig. 6(c,d), this nontrivial case already includes BIPARTIZATION (whose parameterized complexity was open for a long time [49, 60]) and STABLE BIPARTIZATION (whose complexity is resolved first in this paper).

We believe that by obtaining a better understanding of parameterized separation problems, it might be possible to obtain a full dichotomy for parameterized $(H,C,\leq K)$-coloring. However, it is important to note that $(H,C,K)$-coloring behaves very differently, and we cannot hope for a full dichotomy of $(H,C,K)$-coloring at this point. As Fig. 6(e) shows, BIPARTITE INDEPENDENT SET (find disjoint sets $S_1$, $S_2$, each of size $k$, such that there is no edge with one endpoint in $S_1$ and one endpoint in $S_2$) is a special case of $(H,C,K)$-coloring. The parameterized complexity of BIPARTITE INDEPENDENT SET (or equivalently, BICLIQUE in the complement graph) is a longstanding open question. We cannot obtain a dichotomy for $(H,C,K)$-coloring without resolving this question first.

It will be convenient to prove the fixed-parameter tractability result for an even more general problem: in *list $(H,C,\leq K)$-coloring* the input contains a list $L(v) \subseteq V(H)$ for each vertex $v \in V(G)$ and $\theta$ has to satisfy the additional requirement that $\theta(v) \in L(v)$ for every $v \in V(G)$. The main result of the section is the following:

**Theorem 5.1.** *For every fixed H, list $(H,C,\leq K)$-coloring is almost linear-time FPT if $H \setminus C$ is a single edge without loops.*

We start with the introduction of new terminology. Given a graph $G$, a triple $(H,C,K)$ as in the statement of the theorem, and $L : V(G) \to 2^{V(H)}$ associating each vertex of $G$ with the set of allowed vertices of $H$, we say that $\theta$ is an $(H,C,\leq K)$-coloring of $(G,L)$ if $\theta$ is an $(H,C,\leq K)$-coloring of $G$ such that for each $v \in V(G)$, $\theta(v) \in L(v)$. The *exceptional set* of $\theta$ is the set $S$ of all vertices of $G$ that are mapped to $C$ by $\theta$. Since $H \setminus C$ consists of two vertices without loops, $G \setminus S$ is bipartite. Moreover, the size of $S$ is bounded by the parameter $k := \sum_{v \in C} K(v)$. Thus the considered problem is in fact a problem of constrained bipartization.

There is an important detail that makes $(H,C,\leq K)$-coloring different than the bipartization problems of Section 4. The exceptional set of a solution $\theta$ is a bipartization set, but it is not true that there is always a solution whose exceptional set is an inclusionwise minimal bipartization set (see Example 5.3 below). Therefore, we cannot straightforwardly use the approach of Theorem 4.2 and reduce the problem to finding a minimal separator. Nevertheless, we will restrict our attention to solutions where the exceptional set is inclusionwise minimal and argue that treewidth reduction can be performed in a way that preserves all these solutions.

**Definition 5.2.** *An $(H,C,\leq K)$-coloring $\theta$ of $(G,L)$ is* minimal *if there is no $(H,C,\leq K)$-coloring $\theta'$ of $(G,L)$ such that the exceptional set of $\theta'$ is a subset of the exceptional set of $\theta$.*

Observe that if there is an $(H,C,\leq K)$-coloring of $(G,L)$, then there is a minimal $(H,C,\leq K)$-coloring as well.

**Example 5.3.** Let $H$ be a 5-cycle on vertices $\{a,b,c,d,e\}$, let $C = \{c,d,e\}$, and let $K(c) = K(d) = K(e) = 3$ (see Figure 7). Let $G$ be a cycle of length 15 and let $L(v) = \{a,b,c,d,e\}$ for every $v \in V(G)$. Removing any vertex of $G$ makes it bipartite, thus every inclusionwise minimal bipartization set is of size 1. On the other hand, a simple case analysis shows that every solution $\theta$ of $(G,L)$ has at least 3 exceptional vertices. Thus the coloring in Figure 7(b) having 3 exceptional vertices is minimal in the sense of Definition 5.2. The coloring in Figure 7(c) having 9 exceptional vertices is not minimal.

We prove that there is an FPT-computable graph $G^*$ that preserves exceptional sets of all minimal $(H,C,\leq K)$-colorings of $(G,L)$ and whose treewidth is bounded by a function of $k$ (recall that $k = \sum_{v \in C} K(v)$). Similarly to the cases of $\mathcal{G}$-MINCUT and $\mathcal{G}$-BIPARTIZATION, we use this result to transform the given instance of the $(H,C,\leq K)$-coloring problem to an instance with bounded treewidth and then apply Courcelle's Theorem.

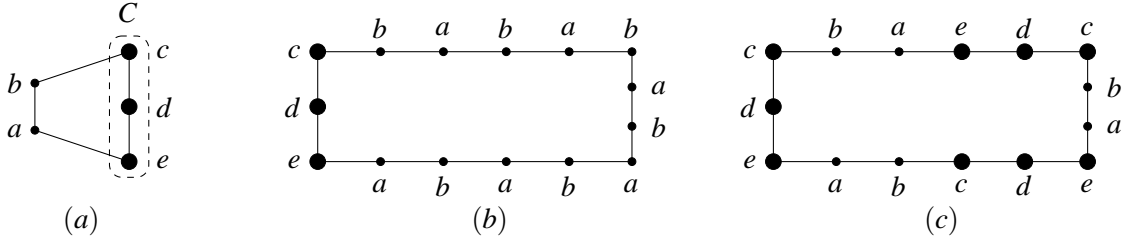

Figure 7: (a) A 5-cycle $H$ with 3 constrained vertices, (b) a minimal coloring of the 15-cycle, (c) a non-minimal coloring of the 15-cycle.

We first show how the treewidth can be reduced in the special case when $G$ is bipartite (Lemma 5.4)), which makes it possible to apply Courcelle's Theorem (Lemma 5.5). It turns out that the general case (when $G$ is nonbipartite) can be easily reduced to the bipartite case: we find a small set $S'$ of vertices whose removal makes the graph bipartite, guess the coloring of these vertices, modify the lists of their neighbors accordingly, and then remove the vertices of $S'$ from the graph.

**Lemma 5.4.** *Assume that $G$ is bipartite. Then there is a linear-time FPT algorithm parameterized by $k = \sum_{v \in V(C)} K(v)$ that finds a set $C''$ such that the treewidth of $\mathrm{torso}(G, C'')$ is at most $f(k, |V(H)|)$ for some function $f$ and the exceptional set of every minimal $(H, C, \leq K)$-coloring of $(G, L)$ is a subset of $C''$.*

*Proof.* The proof is by induction on $k$. For $k = 0$, we can set $C'' = \emptyset$, hence $\mathrm{torso}(G, C'')$ is the empty graph whose treewidth is 0. Assume now that $k > 0$. Denote the vertices of $H \setminus C$ by $b$ and $w$. Let $B$ be the set of all vertices $v \in V(G)$ such that $w \notin L(v)$. Analogously, let $W$ be the set of all vertices $v \in V(G)$ such that $b \notin L(v)$. Let $(B', W')$ be a 2-coloring of $G$ and set $X := (B \cap B') \cup (W \cap W')$ and $Y := (B \cap W') \cup (W \cap B')$ as in Lemma 4.1. If there is no path from $X$ to $Y$, then, by Lemma 4.1, there is a 2-coloring of $G$ where $B$ and $W$ are colored in black and white respectively. In other words, there is a $(H, C, \leq K)$-coloring of $(G, L)$ where each vertex of $G$ is mapped to $b$ and $w$. Consequently, all minimal $(H, C, \leq K)$-colorings of $(G, L)$ have exceptional sets of size 0 and hence $C'' = \emptyset$ as in the case with $k = 0$.

If there is a path connecting $X$ and $Y$, then let us use Lemma 3.9 to compute in FPT-time a set $C'$ such that every minimal set of size at most $k$ separating $X$ and $Y$ in $G$ is a subset of $C'$ and the treewidth of $\mathrm{torso}(G, C')$ is bounded by a function of $k$. Let $P$ be a connected component of $G \setminus C'$ and let $N$ be the subset of $C'$ that consists of all vertices adjacent to the vertices of $P$. Let $\theta$ be an $(H, C, \leq K)$-coloring of $(G[N], L[N])$ where $L[N]$ is the restriction of $L$ to the vertices of $N$. Let $L_\theta$ be the function on $V(P)$ obtained from $L[V(P)]$ by the following operation: for each $v \in V(P)$, remove $u \in L(v)$ from the list of $v$ whenever there is a neighbor $x$ of $v$ in $G$ such that $x \in N$ and $\theta(x)$ is not adjacent to $u$ in $H$. In other words, $L_\theta$ is an updated version of $L$ that allows colors on $V(P)$ so that they are compatible with the mapping of $\theta$ on $N$. Furthermore, let us consider every function $K' : C \to \mathbb{Z}^+$ associating the vertices of $C$ with integers so that $\sum_{v \in C} K'(v) \leq k - 1$. By the induction assumption there is an FPT algorithm parameterized by $k - 1$ that returns a set $C_{P,\theta,K'} \subseteq V(P)$ such that $\mathrm{torso}(P, C_{P,\theta,K'})$ has the treewidth bounded by a $f(k-1, |V(H)|)$ and the exceptional set of any minimal $(H, C, \leq K')$-coloring of $(P, L_\theta)$ is a subset of $C_{P,\theta,K'}$. Let $C_P$ be the union of all possible sets $C_{P,\theta,K'}$. Observe that the number of possible mappings $\theta$ is bounded by a function of $k$ and $|V(H)|$: the vertices of $N$ form a clique in $\mathrm{torso}(G, C')$, hence $|N|$ is bounded by the treewidth of $\mathrm{torso}(G, C')$ plus 1, which is bounded by a function of $k$. Furthermore, the number of possible mappings $K'$ is bounded by a function of $k - 1$ and $|V(H)|$. Therefore by Lemma 2.8, the treewidth of $\mathrm{torso}(P, C_P)$ is bounded by a function of $k$ and $|V(H)|$. Let $C''$ be the union of $C'$ and the sets $C_P$ for all the connected components $P$ of $G \setminus C'$. According to Lemma 2.9, the treewidth of $\mathrm{torso}(G, C'')$ is bounded by $f(k, |V(H)|)$ for an appropriately selected function $f$. (Such function can be defined similarly to function $g$ in the proof of Lemma 2.11). Also, arguing similarly to Lemma 2.11, we can observe that $C''$ can be computed in FPT time parameterized by $k$.

It remains to be shown that the exceptional set $S$ of every minimal $(H,C,\leq K)$-coloring $\theta$ of $(G,L)$ is a subset of $C''$. Since in $G\setminus S$ vertices of $B\setminus S$ are colored in black (i.e., mapped to $b$ by $\theta$) and the vertices of $W\setminus S$ are colored in white (i.e., mapped to $w$ by $\theta$), $S$ separates $X$ and $Y$ according to Lemma 4.1. Therefore, $S$ contains at least one element of $C'$. Consequently, for any connected component $P$ of $G\setminus C'$, $|S\cap V(P)|\leq k-1$. Let $\theta_P$ be the restriction of $\theta$ to the vertices of $P$ and for each vertex $v$ of $C\subseteq V(H)$ define $K'(v)$ as the number of vertices of $P$ mapped to $v$ by $\theta_P$. Let $\theta'$ be the restriction of $\theta$ to the vertices of $C'$ adjacent to $V(P)$. It is not hard to observe that $\theta_P$ is a minimal $(H,C,\leq K')$-coloring of $(P,L_{\theta'})$. In other words, $S\cap V(P) \subseteq C_{P,\theta',K'} \subseteq C_P$. Since each vertex $v$ belongs either to $C'$ or to some $V(P)$, the present lemma follows. □

A standard application of Courcelle's Theorem shows that $(H,C,\leq K)$-coloring is linear-time solvable on bounded treewidth graphs. A possible construction of the required MSO sentence is in the appendix.

**Lemma 5.5.** *For every fixed $H$, $k$, and $w$, $(H,C,\leq K)$-coloring can be solved in linear time, where $w$ is the treewidth of $G$.*

For the proof of Theorem 5.1, we use Lemma 5.4 to reduce treewidth and then apply Lemma 5.5 to solve the problem on the reduced graph. We need additional arguments to handle the coloring of the components of $G\setminus C''$.

*Proof of Theorem 5.1.* First we show that it can be assumed that $G$ is bipartite. Otherwise, we use the FPT algorithm of Kawarabayashi and Reed [44] to find a set $S'$ of at most $k$ vertices whose deletion makes $G$ bipartite (if there is no such set, "NO" can be returned). We branch on the $|V(H)|^{|S'|}$ possible ways of defining $\theta$ on $S'$. For each of these ways we appropriately update the values of $K(v)$ for all $v\in C$. Also, if a vertex $v\in V(G)\setminus S'$ has a neighbor $u\in S'$, then we modify the list of $v$ such that it contains only vertices adjacent to $\theta(u)$ in $H$. It is clear that the original instance has a solution if and only if at least one of the resulting instances has a solution.

As $G$ is bipartite, we can use Lemma 5.4 to obtain the set $C''$. We transform $G$ the following way. Let $P_i$ be a connected component of $G\setminus C''$ having more than one vertex. Let $(X_i,Y_i)$ be the bipartition of $P_i$ (unique due to the connectedness of $P_i$). We replace $P_i$ by two adjacent vertices $x_i$ and $y_i$ such that $x_i$ (resp., $y_i$) is adjacent with the neighborhood of $X_i$ (resp., $Y_i$) in $C''$. We define $L(x_i) = \{b,w\} \cap \bigcap_{v\in X_i} L(v)$, and $L(y_i)$ is defined analogously. Let $G'$ be the graph obtained after performing this operation for every component $P$, and let $L'$ be the resulting list assignment on $G'$. We have $\mathrm{torso}(G,C'') = \mathrm{torso}(G',C'')$ and every component of $G'\setminus C''$ contains at most two vertices. Thus Lemma 2.8 implies that the treewidth of $G'$ is larger than the treewidth of $\mathrm{torso}(G,C'')$ by at most 2 and hence it is bounded by a function of $k$.

We claim that $(G,L)$ has an $(H,C,\leq K)$-coloring if and only if $(G',L')$ has an $(H,C,\leq K)$-coloring, hence we can use Lemma 5.5 to solve the problem. Suppose first that $\theta$ is a minimal $(H,C,\leq K)$-coloring of $(G,L)$. We know that the exceptional set of $\theta$ is contained in $C''$, thus vertices of a component $P$ of $G\setminus C''$ are mapped to $b$ and $w$. Thus if $(X_i,Y_i)$ is the bipartition of $P_i$, then due to $P_i$ being a connected graph, every vertex of $X_i$ is mapped to the same vertex of $H\setminus C$ and similarly for $Y_i$. Thus by mapping $x_i$ and $y_i$ to these vertices in $G'$, we can obtain an $(H,C,\leq K)$-coloring of $(G',L')$.

Suppose now that $(G',L')$ has an $(H,C,\leq K)$-coloring $\theta'$. Again, let $P_i$ be a component of $G\setminus C''$ with bipartition $(X_i,Y_i)$. We can obtain an $(H,C,\leq K)$-coloring $\theta$ of $(G,L)$ by mapping every vertex of $X$ to $\theta'(x_i) \in \{b,w\}$ and every vertex of $Y_i$ to $\theta'(y_i) \in \{b,w\}$. □

# 6 Conclusions

We have presented a general methodology for showing that restricted separation problems can be solved in linear time for every fixed bound $k$ on the size of the separator. This technique allows us to prove fixed-parameter tractability results in an easy way for quite natural problems (such as MINIMUM STABLE $s-t$ CUT) whose parameterized complexity status was open. The connection between separators and bipartization problems, as first developed by Reed et al. [60], makes it possible to use our results for constrained bipartization and certain homomorphism problems.

The results of the paper raise some obvious open questions for future work:

- The treewidth bound of Lemma 2.11 is exponential in $k$, which implies that the running time of the algorithms obtained this way are typically at least double exponential in $k$. Is it possible to improve this dependence on $k$ in the running time to $2^{\text{poly}(k)}$, at least for some basic concrete problems such as MINIMUM STABLE $s-t$ CUT?

- Section 3.3 dealing with CONNECTED $s-t$ CUT shows that our technique can be extended in nontrivial ways to handle certain nonhereditary restrictions. It would be interesting to explore other such extensions.

- Is it possible to extend our techniques to handle global restrictions such as balance requirements (c.f. [24])?

- Is there some way of extending these results (at least partially) to directed graphs?

- Can we introduce treewidth reduction into the MULTICUT algorithm of [54] to obtain more general results? For example, is MULTICUT, parameterized by the size $k$ of the solution, is FPT with the additional restriction that the solution induces an independent set? It would be interesting to see the treewidth reduction of the current paper can be combined with the "random sampling of important separators" technique of [54].

- An obvious goal is to prove an FPT vs. W[1]-hardness dichotomy result for parameterized $(H,C,K)$- and $(H,C,\leq K)$-coloring. For $(H,C,\leq K)$-coloring, this might be doable, but (as discussed in Section 5) for $(H,C,K)$-coloring this would require first understanding the parameterized complexity of BICLIQUE, which is a notorious open problem.

# A  Logical sentences

## A.1  MSO formula in the proof of Theorem 3.1

The part of $\varphi$ describing the separation of $s$ and $t$ is based on the ideas from [28]. The detailed construction is given below. We assume that the two special vertices $s$ and $t$ are labeled in the graph: there is a unary relation $ST = \{s,t\}$.

We construct the formula $\varphi$ as

$$\varphi = \exists C(\text{AtMost}_k(C) \wedge \text{Separates}(C) \wedge \text{Induces}_{\mathcal{G}}(C)),$$

where $\text{AtMost}_k(C)$ is true if and only if $|C| \leq k$, $\text{Separates}(C)$ is true if and only if $C$ separates the vertices of $ST$ in $G^*$, $\text{Induces}_{\mathcal{G}}(C)$ is true if and only $C$ induces a graph of $\mathcal{G}$.

In particular, $\text{AtMost}_k(C)$ states that $C$ does not have $k+1$ mutually non-equal elements: this can be implemented as

$$\forall c_1, \ldots, \forall c_{k+1} \bigvee_{1 \leq i,j \leq k+1} (c_i = c_j).$$

Formula $\text{Separates}(C)$ is a slightly modified formula $\text{uvmc}(X)$ from [28] that looks as follows:

$$\forall s \forall t \big((ST(s) \wedge ST(t) \wedge \neg(s = t))\big) \rightarrow \big(\neg C(s) \wedge \neg C(t) \wedge \forall Z(\text{Connects}(Z,s,t) \rightarrow \exists v(C(v) \wedge Z(v)))\big),$$

where $\text{Connects}(Z,s,t)$ is true if and only if in the modeling graph there is a path from $s$ and $t$ all vertices of which belong to $Z$. For definition of the predicate Connects, see Definition 3.1 in [28]

To construct $\text{Induces}_{\mathcal{G}}(C)$, we explore all possible graphs having at most $k$ vertices and for each of these graphs we check whether it belongs to $\mathcal{G}$. Since the number of graphs being explored depends on $k$ and $\mathcal{G}$ is a decidable class, we can compile the set $\{G'_1, \ldots, G'_r\}$ of all graphs of at most $k$ vertices that belong to $\mathcal{G}$ in time depending only on $k$. Let $k_1, \ldots k_r$ be the respective numbers of vertices of $G'_1, \ldots G'_r$. Then $\text{Induces}_{\mathcal{G}}(C) = \text{Induces}_1(C) \vee \cdots \vee \text{Induces}_r(C)$, where $\text{Induces}_i(C)$ states that $C$ induces $G'_i$. To define $\text{Induces}_i$, let $v_1, \ldots v_{k_i}$ be the set of vertices of $G'_i$ and define $\text{Adjacency}(c_1, \ldots, c_{k_i})$ as the conjunction of all $E(c_x, c_y)$ such that $v_x$ and $v_y$ are adjacent in $G'_i$. Then

$$\text{Induces}_i(C) = \text{AtMost}_{k_i}(C) \wedge \exists c_1 \ldots \exists c_{k_i} \bigg( \bigwedge_{1 \leq j \leq k_i} C(c_j) \wedge \bigwedge_{1 \leq x,y \leq k_i} (c_x \neq c_y) \wedge \text{Adjacency}(c_1, \ldots, c_{k_i}) \bigg).$$

Let us now verify that indeed $G_1 \models \varphi$ if and only if $(G^*, s, t, k)$ is a 'YES' instance of the $\mathcal{G}$-MINCUT problem. Assume first the latter and let $S$ be an $s-t$ separator of size at most $k$ such that $G^*[S] \in \mathcal{G}$. Let us observe that all the three main conjuncts of $\varphi$ quantified by $C$ are satisfied when $S$ is substituted instead $C$. That $\text{AtMost}_k(S)$ is true immediately follows from the pigeonhole principle: if we take $k+1$ elements out of a set of at most $k$ elements, at least 2 of them must be equal. To show that $\text{Separates}(S)$ is true w.r.t. $G_1$, we draw the following line of implications. Set $S$ separates $s$ and $t$ in $G^*$, hence the set of vertices of every path from $s$ to $t$ intersects with $S$, hence every set $Z$ including as a subset a set of vertices of a path from $s$ to $t$ intersects with $S$. Formally written, the last statement can be expressed as follows $\forall Z(\text{Connects}(Z,s,t) \rightarrow \exists v(S(v) \wedge Z(v)))$, but this (together with the fact that $S$ is disjoint with $\{s,t\}$) is the right-hand part of the main implication of $\text{Separates}(S)$, hence $\text{Separates}(S)$ is true. To verify that $\text{Induces}_{\mathcal{G}}(S)$ is true w.r.t. $G_1$, let $G'_i \in \mathcal{G}$ be the graph isomorphic to $G^*[S]$ and observe that $\text{Induces}_i(S)$ is true by construction.

For the opposite direction assume that $G_1 \models \varphi$. It follows that there is a set of vertices $C$ such that $\text{AtMost}_k(C)$, $\text{Separates}(C)$, and $\text{Induces}_{\mathcal{G}}(C)$ are all true. Consequently, $|C| \leq k$. Indeed otherwise, we can select $k+1$ *distinct* elements of $C$ that falsify $\text{AtMost}_k(C)$. It also follows that $C$ is disjoint with $\{s,t\}$ and separates $s$ from $t$ in $G^*$. Indeed $s$ and $t$ satisfy the left part of the main implication of Separates($C$), hence the right part of it must be satisfied as well. It immediately implies that $C$ is disjoint with $s$ and $t$. If we assume that $C$ does not separate $s$ and $t$ then there is a path $P$ from $s$ to $t$ avoiding $C$. Let $Z = V(P)$. Then Connects($V(P), s, t$) is true while $\exists v(C(v) \wedge Z(v))$ is false, falsifying the last conjunct of the right part of the main implication, a contradiction. Finally, it follows from Induces$_G(C)$ that Induces$_i(C)$ is true for some $i$. By construction, this means that $G^*[C]$ is isomorphic to $G'_i \in \mathcal{G}$. Thus $(G^*, s, t, k)$ is a 'YES' instance of the $\mathcal{G}$-MINCUT problem.

## A.2 MSO formula in the proof of Lemma 5.5

Let $(G, L)$ be an instance of the $(H, C, \leq K)$-coloring. For each $x \in V(H)$, let $L_x$ be the subset of $V(G)$ consisting of all vertices $v$ such that $x \in L(v)$. Denote the vertices of $H$ by $x_1, \ldots, x_r$ and let $G_1 = (V(G), E(G), L_{x_1}, \ldots, L_{x_r})$ be a labeled graph. We construct a formula $\varphi$ such that $G_1 \models \varphi$ if and only if there is a $(H, C, \leq K)$-coloring of $(G, L)$.

The formula $\varphi$ is defined as

$$\exists V_1 \exists V_2 \ldots \exists V_r \bigg( \bigwedge_{\substack{1 \leq i \leq r \\ x_i \in C}} \text{AtMost}_{K(x_i)}(V_i) \wedge \text{partition}(V_1, \ldots, V_r) \wedge \bigwedge_{1 \leq i \leq r} \text{subset}(V_i, L_{x_i})$$

$$\wedge \bigwedge_{\substack{x_i, x_j \in V(H) \\ x_i x_j \notin E(H)}} \forall v, u((V_i(v) \wedge V_i(u)) \to \neg E(v, u)) \bigg),$$

where

$$\text{partition}(V_1, \ldots, V_r) := \bigg( \bigwedge_{1 \leq i < j \leq r} \text{disjoint}(V_i, V_i) \bigg) \wedge \bigg( \forall v \bigvee_{1 \leq i \leq r} V_i(v) \bigg)$$

expresses that $(V_1, \ldots, V_r)$ is a partition of $V(G)$ and

$$\text{subset}(A, B) := \forall v(A(v) \to B(v))$$

is true if and only if $A \subseteq B$.

It is not hard to see that $G_1 \models \varphi$ if and only if for some choice of $V_1, \ldots, V_r$ there is an $(H, C, \leq K)$-coloring $\theta$ of $(G, L)$ defined by $\theta(v) = x_i$ if and only if $v \in V_i$. Note furthermore that the length of $\varphi$ depends only on $k$ and $H$.